\documentclass[journal,twocolumn]{IEEEtran}

\usepackage{cite}
\usepackage{amsmath,amsfonts,amsxtra,amssymb,latexsym,amscd,amsthm,mathrsfs,bm}
\usepackage{graphicx}
\graphicspath{{./figures/}}
\usepackage{hyperref}%%% this is giving me warning
\usepackage{xcolor}
\usepackage{algorithm}
\usepackage{algpseudocode}

\usepackage[small]{caption} 
\usepackage{epstopdf}% convert eps figure automatically to pdf for pdflatex
\usepackage{comment}
\usepackage{mathtools}
\newcommand{\RomanNumeralCaps}[1]
    {\MakeUppercase{\romannumeral #1}}

\makeatletter
\def\BState{\State\hskip-\ALG@thistlm}

\begin{document}

\title{{IM-based Pilot-assisted Channel Estimation for FTN Signaling HF Communications}}

\author{\IEEEauthorblockN{Simin Keykhosravi and Ebrahim Bedeer}\\
\IEEEauthorblockA{\textit{Department of Electrical and Computer Engineering, University of Saskatchewan, Saskatoon, Canada} \\
Emails:{\{sik904, e.bedeer\}}@usask.ca}}

\maketitle
\begin{abstract}
This paper {investigates} doubly-selective (i.e., time- and frequency-selective) channel estimation in faster-than-Nyquist (FTN) signaling HF communications. 
{In particular, we propose a novel IM-based channel estimation algorithm for FTN signaling HF communications including pilot sequence placement (PSP) and pilot sequence location identification (PSLI) algorithms.} {At the transmitter, we propose the PSP algorithm that utilizes the locations of pilot sequences to carry additional information bits, thereby improving the SE of HF communications.} 
{HF channels have two non-zero independent fading paths with specific fixed delay spread and frequency spread characteristics as outlined in the Union Radio communication Sector (ITU-R) F.1487 and F.520. Having said that, based on the aforementioned properties of the HF channels and the favorable auto-correlation characteristics of the optimal pilot sequence, we propose a novel PSLI algorithm that effectively identifies the pilot sequence location within a given frame at the receiver.}
This is achieved by showing that the square of the absolute value of the cross-correlation between the received symbols and the pilot sequence consists of a scaled version of the square of the absolute value of the auto-correlation of the pilot sequence weighted by the gain of the corresponding HF channel path. 
Simulation results show very low pilot sequence location identification errors for HF channels.
 {Our simulation results show a 6 dB improvement in the MSE of the channel estimation as well as about 3.5 dB BER improvement of FTN signaling along with an enhancement in SE compared to the method in \cite{ishihara2017iterative}. We also achieved an enhancement in SE compared to the work in \cite{keykhosravi2023pilot} while maintaining comparable MSE of the channel estimation and BER performance. }
\end{abstract}


\section{Introduction}

Diverse applications of wireless communications %necessitate 
{require} transmitting information over long distances in rural and remote areas. High-frequency (HF) communications represent an effective method of communicating with long-distance users in remote or isolated locations without access to other wireless services \cite{wang2018hf}.
The frequencies between 3 MHz and 30 MHz are denoted as the HF band, which represents a major part of the shortwave band. Refraction of radio waves from the ionosphere back to earth known as skywave propagation facilitates long-distance communications\cite{muresan2015ionospheric}. Usually, the allocated bandwidth for the HF communication channels is limited to 3 kHz \cite{navarro2004error}. Hence, HF communications have a bottleneck because of its small bandwidth and limited data rate. Having limited bandwidth in HF places a {significant} demand on enhancing the spectral efficiency (SE) for long-distance communications. 

{Introduced in the 1970s, faster-than-Nyquist (FTN) signaling has increasingly gained researchers' interest as a potential candidate to enhance the SE for a given bandwidth and energy per bit} \cite{anderson2013faster, li2020beyond}. The conventional Nyquist transmission employing $T$-orthogonal pulses specifies a maximum rate of $1/T$ to have inter-symbol interference (ISI)-free communications over frequency flat channels known as the Nyquist limit \cite{john2008digital}. FTN signaling, on the other hand, %entails 
{involves} the transmission of data symbols utilizing $T$-orthogonal pulses at symbol intervals of $\tau T$, where {$0 < \tau <1$} is the FTN signaling acceleration parameter. Exceeding the Nyquist limit results in a higher transmission rate, albeit at the expense of inevitable ISI at the receiver. 

{ The detection of FTN signaling is addressed in several works, e.g.,{\cite{liveris2003exploiting, prlja2008receivers,bedeer2017reduced, hong2016performance, sugiura2013frequency,  shi2017frequency, bedeer2017very,ibrahim2021novel,li2020beyond,hirano2014tdm, ishihara2017iterative, wu2017hybrid,keykhosravi2023pilot}.} The ISI, introduced by the uncoded transmission of FTN signaling, has been demonstrated to exhibit a trellis structure \cite{liveris2003exploiting}. Therefore, in order to mitigate the effects of trellis-structured ISI, one can utilize a trellis decoder for data detection. Due to the high computational complexity associated with the optimal FTN signaling detection for small values of $\tau$ even in the presence of additive white Gaussian noise (AWGN), various studies have proposed reduced trellis or reduced tree search methods to approximate the optimal solution \cite{liveris2003exploiting, prlja2008receivers, bedeer2017reduced}. 
Tree search and trellis-based equalizers become challenging to employ in scenarios with high modulation orders due to their excessive complexity.} 

{Early studies of FTN signaling over AWGN channel employ time-domain optimal detectors to address the ISI \cite{prlja2008receivers}.
In scenarios with high values of $\tau$, methods such as frequency domain equalization (FDE) \cite{sugiura2013frequency,hong2016performance,shi2017frequency} or symbol-by-symbol detection \cite{bedeer2017very} are used to design low complexity detectors for FTN signaling.
For high-order and ultra-high order modulations where FTN signaling detection becomes even more challenging, precoding techniques \cite{li2020beyond} and optimization methods based on the alternating directions multiplier method (ADMM) \cite{ibrahim2021novel} prove effective in reducing the computational complexity.
}

The channel estimation of FTN signaling across frequency-selective channels has been studied in a few papers \cite{hirano2014tdm, wu2017hybrid, ishihara2017iterative,keykhosravi2023pilot}. 
In order to estimate the channel in \cite{hirano2014tdm}, a Nyquist-based pilot sequence was {inserted} before the data block. In contrast to \cite{hirano2014tdm}, in \cite{wu2017hybrid} an FTN signaling-based pilot sequence was utilized for time-domain joint channel estimation and FTN signaling detection. In \cite{ishihara2017iterative}, a low-complexity joint channel estimation and data detection for FTN signaling was developed in the frequency domain to {reduce} the complexity. In particular, a guard block with the same length as the pilot sequence is introduced ahead of the transmitted data within the frame to prevent the ISI from affecting the FTN signaling pilot sequence. Similarly, a cyclic prefix of the same length as the pilot sequence and the guard block was {inserted} at the end of the transmitted data in the frame. Such overhead which is four times the length of the pilot sequence within the frame will eventually reduce the SE in \cite{ishihara2017iterative}. 

{
In most works on the detection of FTN signaling over frequency-selective channels,
the channel is regarded as quasi-static. While \cite{hirano2014tdm,wu2017hybrid,ishihara2017iterative} assume a quasi-static channel, a
time-selective channel was considered in \cite{shi2017frequency}. In \cite{shi2017frequency}, a technique for frequency domain joint channel estimation and FTN signaling detection is proposed, relying on the implementation of the message-passing algorithm.
Additionally, the paper in \cite{keykhosravi2023pilot} proposed a novel channel estimation technique for FTN signaling over doubly-selective channels based on the least sum of squared errors (LSSE) approach which takes the ISI caused by both FTN signaling and the frequency-selective channel into account. The optimal pilot sequence that minimizes the mean square error (MSE) of the channel estimation is found in \cite{keykhosravi2023pilot}. For the time-selective nature of the channel, a low-complexity linear interpolation is employed in \cite{keykhosravi2023pilot} to track complex channel coefficients at data symbol locations within the frame. In contrast to \cite{ishihara2017iterative}, the implemented frame structure in \cite{keykhosravi2023pilot} is {more SE efficient} since it {eliminates} the necessity for an extra cyclic prefix, thereby {reducing the overhead}.}

{In both Nyquist and FTN signaling studies, inserting a pilot sequence for the channel estimation purpose results in a SE reduction. Thus, there is a need to use methods to further improve the SE. {Due to the restricted bandwidth in HF communication, it is crucial to improve the SE.} 
The index modulation (IM) technique has drawn the interest of researchers owing to its capability of enhancing the transmission rate by activating a selection of indices to carry extra information \cite{mao2018novel}. } {IM is also employed in the frequency domain to convey additional bits over the activated subcarrier indices in conjunction with the $M$-ary modulation in \cite{ma2020joint}. The work in \cite{ma2020joint} proposes a low-complexity frequency-domain joint channel estimation and equalization algorithm for spectrally efficient frequency division multiplexing (SEFDM) based on IM systems communicating over frequency selective fading channels.}

{Inspired by IM techniques, in this paper, we propose a novel IM-based channel estimation algorithm including pilot sequence placement (PSP) and pilot sequence location identification (PSLI) algorithms. The main contributions of this paper can be outlined as follows:}

{
\begin{itemize}
    \item 
At the transmitter, we propose the PSP algorithm, which makes use of the pilot sequence locations to transmit extra information bits. 
That is to say, the location of the pilot sequence in each frame is not fixed and it is determined based on the information bits. 
\item 
{At the receiver, we propose a novel PSLI algorithm to detect the incorporated information bits from the pilot sequence location. Pilot sequence location identification is achieved by taking the inherent characteristics of the HF channel as well as the autocorrelation properties of the optimal pilot sequence that minimizes the mean square error (MSE) of the channel estimation into account. } 
On one hand, {based on} our analysis of the comprehensive array of test HF channels introduced in the International Telecommunication Union Radio Communication Sector (ITU-R) F.1487 document, we make use of the fact that each HF channel comprises two non-zero independent fading paths with specific fixed delay spread and frequency spread attributes.
On the other hand, our observation of the square of the absolute value of the autocorrelation exhibited by the optimal pilot sequence reveals favorable local characteristics. In particular, we show that for the received symbols corresponding to the pilot sequence, the square of the absolute value of the cross-correlation between the received symbols and the pilot sequence contains a scaled version of the square of the absolute value of the auto-correlation of the pilot sequence weighted by the gain corresponding to the HF channel path. We exploit these identified {characteristics to find the location of} the pilot sequence within a frame. 
\item 
 The simulation results of the proposed IM-based channel estimation algorithm for FTN signaling demonstrate exceptionally low pilot sequence location identification error (less than $2 \times 10^{-6}$ and $10^{-3}$ for $E_\text{b}/N_0$ values higher than 6 dB and 12 dB for time-invariant and time-variant HF Poor channel, respectively, for $\tau=0.84$). Our simulation results indicate achieving comparable performance in terms of MSE and BER of the channel estimation while having an improvement in SE when compared to the work in \cite{keykhosravi2023pilot} over frequency-selective channels. Simulation results also demonstrate an enhancement in SE along with noteworthy improvement (about 6 dB) in the MSE of the channel estimation as well as (about 3.5 dB) in BER of FTN signaling compared to the approach introduced in \cite{ishihara2017iterative} when dealing with frequency-selective channels. 
\end{itemize}
}
The {following} sections of this paper are structured as follows. 
Section~\RomanNumeralCaps{2} {introduces} the system model of the proposed IM-based HF channel estimation FTN signaling system and {familiarizes} the reader with the adopted frame structure. Section ~\RomanNumeralCaps{3} proposes a PSP algorithm to place the pilot sequence in each frame at the transmitter in a way that facilitates the detection in the receiver. Section ~\RomanNumeralCaps{4} elucidates the proposed PSLI algorithm specifically tailored for the HF channel. Section ~\RomanNumeralCaps{4} {presents} the simulation results, emphasizing the outcomes achieved through the conducted simulations. Section ~\RomanNumeralCaps{5} serves as the final section, offering concise conclusions derived from the preceding sections of the paper.

In this paper, column vectors are denoted by boldface lowercase letters, such as $\mathbf{x}$, while scalars are represented by lightface lowercase letters, such as $x$. The complex conjugate of a complex number $x$ is indicated by $x^\ast$, and the convolution operator between two variables $x$ and $y$ is denoted as $x \ast y$. The expectation operator for a vector $\mathbf{x}$ is symbolized as $\text{E}(\mathbf{x})$, and we use $\backslash$ to exclude an element from a set. 
We use $\mathbf{x}^\text{H}$ to denote the complex conjugate transpose (Hermitian transpose) operations of vector $\mathbf{c}$.
The ceil operator represented as $\lceil{x}\rceil$ rounds the real value $x$ up to the nearest integer greater than or equal to $x$. The floor operator represented as $\lfloor{x}\rfloor$ rounds the real value $x$ down to the nearest integer less than or equal to $x$. The boldface uppercase letters, e.g., $\mathbf{X}$, are employed to represent matrices. Additionally, the $\mathbb{C}^{N\times M}$ encompasses all complex matrices of complex matrices of size ${N\times M}$. To represent the inverse, transpose, and complex conjugate transpose operations of a matrix $\mathbf{X}$, we employ $\mathbf{X}^{-1}$, $\mathbf{X}^\text{T}$, and $\mathbf{X}^\text{H}$, respectively. Finally, the cardinality of a set $\mathbb{P}$ is defined by $\text{Card}\{\mathbb{P}\}$.

\begin{figure*}[t]
\centering
\includegraphics[width=.95\textwidth]{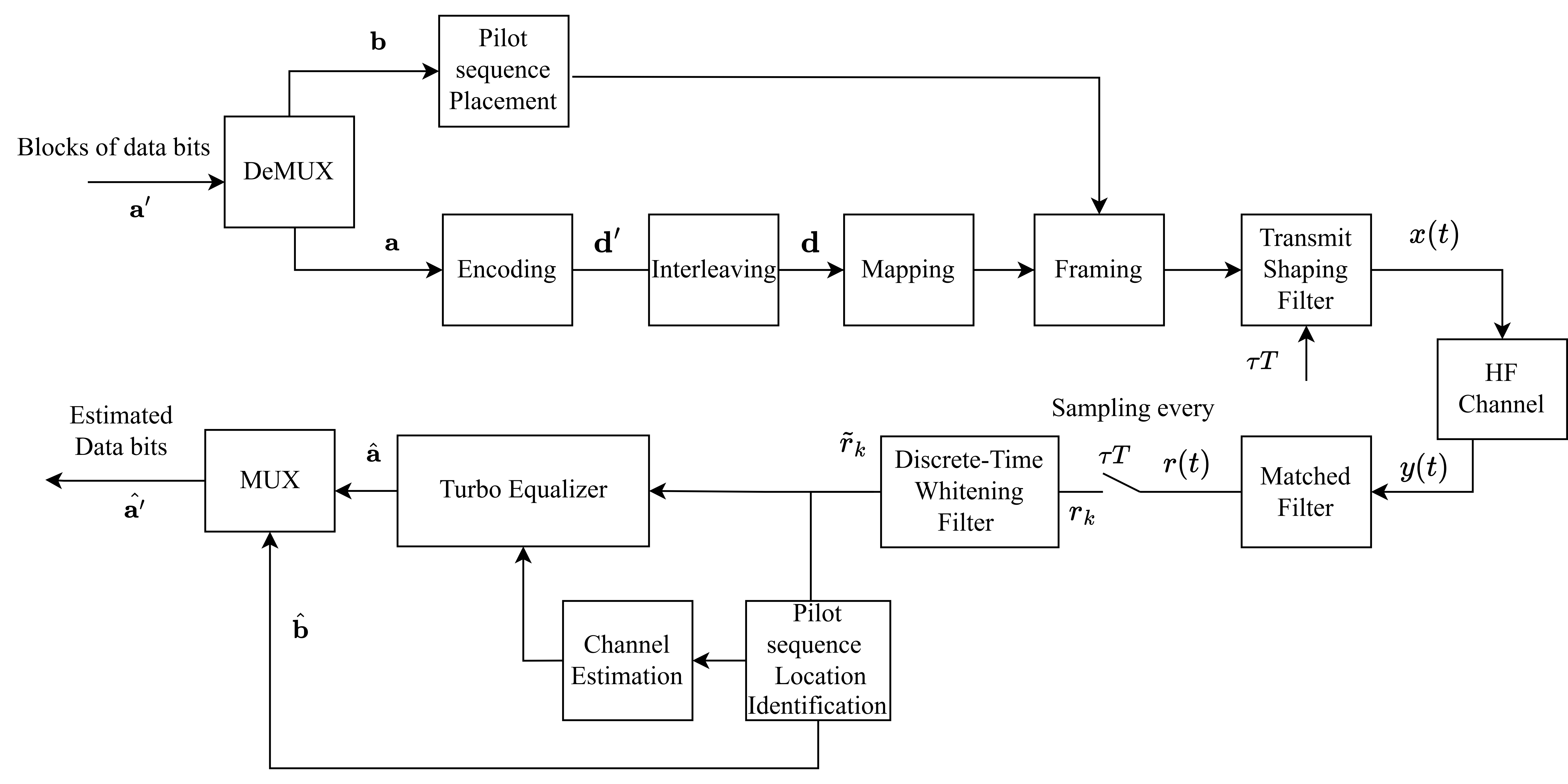}
\caption{Block diagram of a communication system employing IM-based HF channel estimation FTN signaling system.}
\label{fig:66}
\end{figure*}

\section{System Model}
In this paper, we employ an IM-based HF channel estimation FTN signaling system for long-distance communications as depicted in Fig. \ref{fig:66}. Each block of {independent and identically distributed (i.i.d.)} $N_\text{d}'$ data bits, denoted as $\textbf{a}'= {[a, b]}^\text{T}$, {consists of} a block of $N_\text{d}$ regular information bits $\textbf{a}={[a_0, a_1, ..., a_{N_\text{d}-1}]}^\text{T}$ along with $N_b$ additional bits $\textbf{b}={[b_0, b_1, ..., b_{N_b-1}]}^\text{T}$ that will be used to specify the pilot sequence location in the frame.         

A block of $N_\text{d}$ regular information bits, $\textbf{a}$, to be transmitted in a given frame is first encoded to $\textbf{d}'$ and then is shuffled by an interleaver to $\textbf{d}$. Each ${\log_2}M$ interleaver's output bits are then modulated onto a complex data symbol belonging to an $M$-ary quadrature amplitude modulation ($M$-QAM) constellation set $\mathcal{S}_\text{d}$.  A frame of length $N = N_\text{p} + {{N_\text{s}}}$ is constructed by inserting known pilot sequence $\textbf{p} = [p_0, p_1, ...,p_{N_\text{p}-1}]^{\text{T}}$ of length $N_\text{p}$ in the location determined by the $N_b$ information bits to assist the receiver in the channel estimation. {Please note that $N_\text{s} = \frac{N_\text{d}}{{\log_2}M}$ is the number of symbols in a frame.
As mentioned earlier, the location of the pilot sequence, $n_{s_p}^{(i)}$, is not fixed in a given frame $i$ and conveys additional $N_b$ data bits. At the transmitter, the pilot sequence could be located at any location belonging to $n_{s_p}^{(i)}=\{0,1, ..., N_\text{s}\}$, and hence, one has $N_\text{s}+1$ choices to place the pilot in a frame, which is equivalent to carrying $\lfloor {\log_2}{(N_s+1)} \rfloor$ extra bits.  
It is worth mentioning that if the $ {\log_2}{(N_s+1)}$ is not an integer number, the number of bits carried by the location of pilot sequence is $\lfloor {\log_2}{(N_s+1)} \rfloor< {{\log_2}{(N_s+1)}}$ and this is equivalent to $2^{\lfloor {\log_2}{(N_s+1)} \rfloor}$ (which is less than ${N_s+1}$) choices for the pilot sequence to be located to construct the frame and thus the pilot sequence location is $n_{s_p}^{(i)}=\{0,1, ..., 2^{\lfloor {\log_2}{(N_s+1)} \rfloor}-1\}$. }
In Section ~\RomanNumeralCaps{3}, we describe the proposed PSP algorithm for transmitting the $N_b$ additional bits through the pilot sequence location in each frame at the transmitter and a low complexity pilot sequence identification algorithm for identification of the pilot sequence location and thus detecting the $N_b$ bits for each frame at the receiver in Section ~\RomanNumeralCaps{4}.

\begin{figure*}[t]
\centering
\hspace{2cm} % Adjust the value as needed
\includegraphics[width=0.85\textwidth]{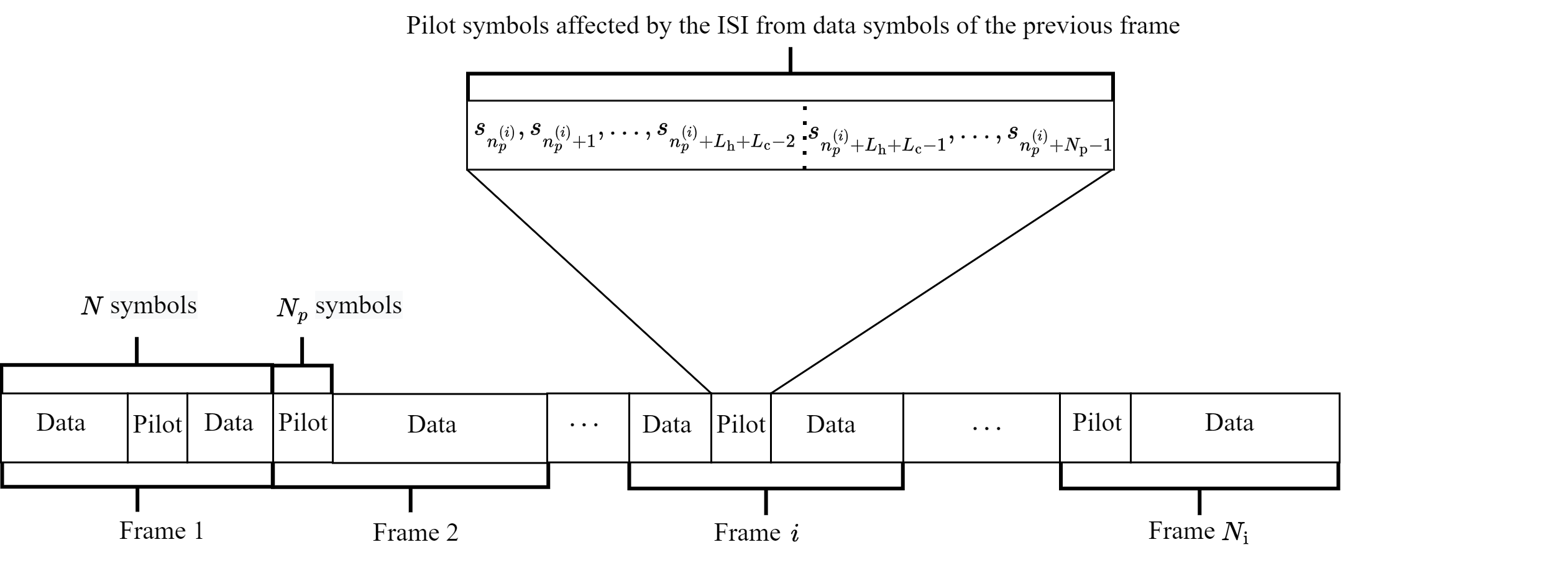}
\caption{The frame structure with dynamic pilot sequence location in frames employed in the paper. }
\label{fig:2}
\end{figure*}

The pilot sequence and data symbols within a frame are transmitted every symbol interval $\tau T$ through a $T$-orthogonal root raised cosine (RRC) pulse, $g(t)$, where $g(t)$ is a unit energy pulse, i.e., $\int_{-\infty}^{\infty} |g(t)|^2 \,dt = 1$. Having said that, the baseband transmitted signal can be expressed as 
\begin{IEEEeqnarray}{rcl}\label{equation1}
x(t) &{}={}& \sum \limits_{i=1}^{N_i} \sum \limits_{n_p^{(i)} \in \Omega_\text{p}^{(i)}}  {s_{n_p^{(i)}}  g(t-n_p^{(i)} \tau T)}  \nonumber \\
& &{}+ \sum \limits_{i=1}^{N_i} \sum \limits_{n_d^{(i)} \in \Omega_\text{d}^{(i)}}  {s_{n_d^{(i)}}  g(t-n_d^{(i)} \tau T)}, \IEEEeqnarraynumspace
\end{IEEEeqnarray}
where $N_i$ is the total number of frames, $\Omega_\text{p}^{(i)} = \{n_{s_p}^{(i)}, n_{s_p}^{(i)}+1, \ldots, n_{s_p}^{(i)}+N_p-1 \}$ is the set of pilot sequence symbols' locations within the $i$th frame starting from the $n_{s_p}^{(i)}$th symbol of pilot sequence in the frame $i$ as can be seen in Fig. \ref{fig:2}. That is to say for the frame $i$, we have $[s_{n_{s_p}^{(i)}}, s_{n_{s_p}^{(i)}+1}, \ldots, s_{n_{s_p}^{(i)}+N_p-1}]^{\text{T}} = [p_0, p_1, ...,p_{N_\text{p}-1}]^{\text{T}}$.
We define $\Omega_\text{d}^{(i)} = \Omega \backslash \Omega_\text{p}^{(i)}$ as the set of data symbols' locations within the $i$th frame where  $n_d^{(i)} \in \Omega_\text{d}^{(i)}$, and ${s_{n_d^{(i)}}}$ is the data symbol within the frame $i$. We have {$\Omega = \{0, 1, , \ldots, N-1 \}$}. 

Considering the transmission of FTN signaling over an HF doubly-selective fading channel denoted as $c(t,\phi)$, the received signal can be expressed as 
\begin{IEEEeqnarray}{rcl}\label{equation2}
y(t) &{}={}& x(t)\ast c(t,\phi) + n(t),
\end{IEEEeqnarray}
where $n(t)$ is the zero-mean AWGN with double-sided power spectral density $N_0/2$. The received signal, $y(t)$, is passed through a filter matched to $g(t)$, $g^\ast(-t) = g(t)$, and the output of the matched filter is represented as
\begin{IEEEeqnarray}{rcl}\label{equation4}
r(t) &{}={}& x(t)\ast c(t,\phi)\ast g(t)+z(t),
\end{IEEEeqnarray}
where $z(t) = n(t)\ast g(t)$. Substituting {$x(t)$} from (\ref{equation1}), the received signal in (\ref{equation4}) can be re-expressed as
\begin{IEEEeqnarray}{rcl}\label{equation5}
r(t) &{}={}&  \sum \limits_{i=1}^{N_i}  \sum \limits_{n_p^{(i)} \in \Omega_\text{p}^{(i)}}  {s_{n_p^{(i)}}  h(t-n_p^{(i)} \tau T)} \ast c(t,\phi) \nonumber \\
& &{}+ \sum \limits_{i=1}^{N_i} \sum \limits_{n_d^{(i)} \in \Omega_\text{d}^{(i)}}  {s_{n_d^{(i)}}  h(t-n_d^{(i)} \tau T)} \ast c(t,\phi)\nonumber \\
& &{}+ z(t),
\end{IEEEeqnarray}
where $h(t) = g(t)\ast g(t)$ is the RC pulse. While the ISI length of the FTN signaling is theoretically infinite, we consider the fact {that %predominantly 
the} power of the raised cosine pulse, $h(t)$, is concentrated in and around its main lobe. Consequently, we limit the length of the raised cosine pulse, $h(t)$, to $2L_\text{h}+1$. In other words, we truncate the pulse to have $2L_\text{h}+1$ effective taps, which can be represented by the vector $\mathbf{h} = [h_{-L_\text{h}},h_{L_\text{h}+1},\hdots,h_{L_\text{h}}]^\text{T}$. The parameter $L_\text{h}$ depends on the FTN signaling packing ratio $\tau$ and the roll-off factor of the RRC pulse $g(t)$.

Being widely employed in the simulation of the HF ionospheric channel, the Watterson model \cite{cavers2006mobile} defines a tapped delay line, with $L_\text{c}+1$ taps, each one models an ionospheric propagation path. Hence, we utilize the widely adopted tapped delay line channel model to represent a doubly-selective channel. In this model, the impulse response of the channel, denoted as $c(t,\phi)$, consists of a total of $L_\text{c}+1$ channel taps, namely $c_l(t)$ where $l$ ranges from 0 to ${L_\text{c}}$. Each channel path $c_l(t)$ within this model corresponds to a specific delay $\phi_l (t)$. The output of the matched filter in (\ref{equation5}) is sampled at intervals of $\tau T$, which can be mathematically represented as 
\begin{IEEEeqnarray}{rcl}\label{equation6}
r_k &{}={}&  \sum \limits_{i=1}^{N_i}  \sum \limits_{n_p^{(i)} \in \Omega_\text{p}^{(i)}}  {s_{n_p^{(i)}} \sum\limits_{l=0}^{L_\text{c}} c_{k,l} h_{k-l-n_p^{(i)}} } \nonumber \\
& &{}+ \sum \limits_{i=1}^{N_i}  \sum \limits_{n_d^{(i)} \in \Omega_\text{d}^{(i)}} {s_{n_d^{(i)}}  \sum\limits_{l=0}^{L_\text{c}} c_{k,l} h_{k-l-n_d^{(i)} }} + z_k,
\end{IEEEeqnarray}

In order to decorrelate the colored noise samples $z_k$, the sampled signal in (\ref{equation6}) passes through a discrete-time whitening filter, $\textbf{v}=[v_0, v_1, ..., v_{L_\text{h}-1}]^\text{T}$ of length $L_\text{h}$ as explained in \cite{keykhosravi2023pilot}.
The $k$th output sample of the whitening filter can be written as
\begin{IEEEeqnarray}{rcl}\label{equation7}
\tilde{r}_k &{}={}&  \sum \limits_{i=1}^{N_i}  \sum \limits_{n_p^{(i)} \in \Omega_\text{p}^{(i)}}  {s_{n_p^{(i)}} \sum\limits_{l=0}^{L_\text{c}} c_{k,l} v_{k-l-n_p^{(i)}} } \nonumber \\
& &{}+  \sum \limits_{i=1}^{N_i}\sum \limits_{n_d^{(i)} \in \Omega_\text{d}^{(i)}} {s_{n_d^{(i)}}  \sum\limits_{l=0}^{L_\text{c}} c_{k,l} v_{k-l-n_d^{(i)} }} + w_k, 
\end{IEEEeqnarray}
where $w_k$ is the zero-mean white Gaussian noise with variance $\sigma_n^2$.

The output of the whitening filter, as introduced in (\ref{equation7}), is then fed into the PSLI block to identify the pilot sequence's location within a given frame. Once the pilot sequence's location is determined, the receiver {employs} %proceeds with employing 
the least sum of squared errors (LSSE) based approach, as proposed in \cite{keykhosravi2023pilot}, to estimate the complex channel coefficients at the %designated 
pilot sequence locations within the frame. The turbo equalizer receives the samples output of the whitening filter, alongside the estimated channel impulse response as an input from the channel estimation block, and outputs the estimated data blocks $\hat{\textbf{a}}$. Finally, the estimated regular bits output of the turbo equalizer, along with the bits representing the pilot sequence's location in the frame, are multiplexed in the MuX block to generate the estimated data block $\hat{\textbf{a}'}$.
%%%%%%%%%%%%%%%%%%%%%%%%
\begin{figure}[t]
\centering
\includegraphics[width=8.5cm,height=6cm]{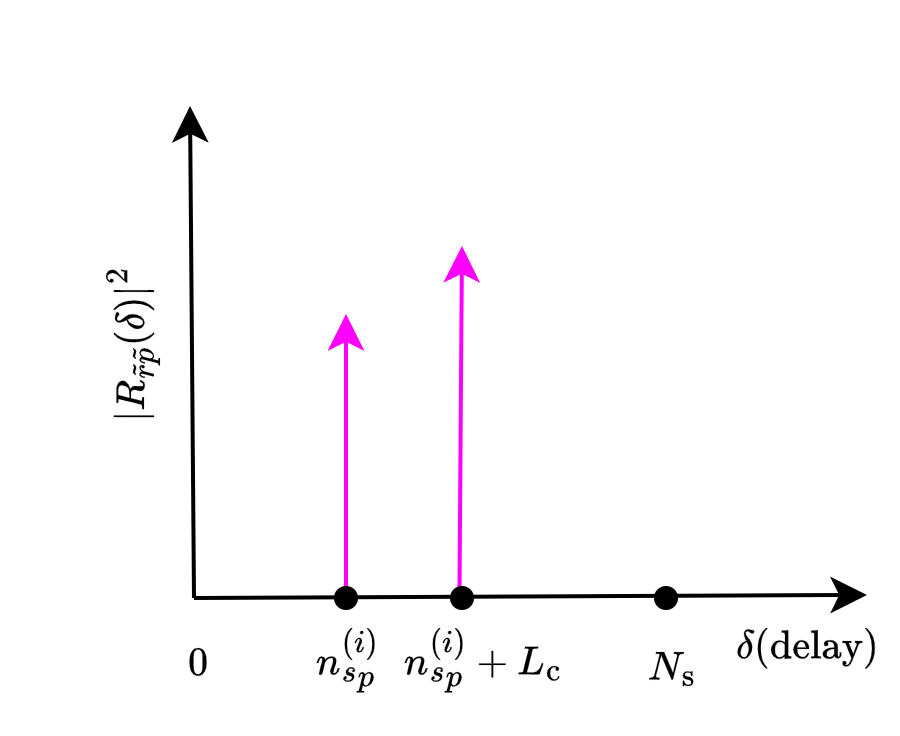} %
\caption{Example for the square of the absolute value of the cross-correlation of the received signal and the pilot sequence.}
\label{fig:1111}
\end{figure}
%%%%%%%%%%%%%%%%%%%%%%%%
\begin{figure*}[t]
\centering
\includegraphics[width=0.85\textwidth]{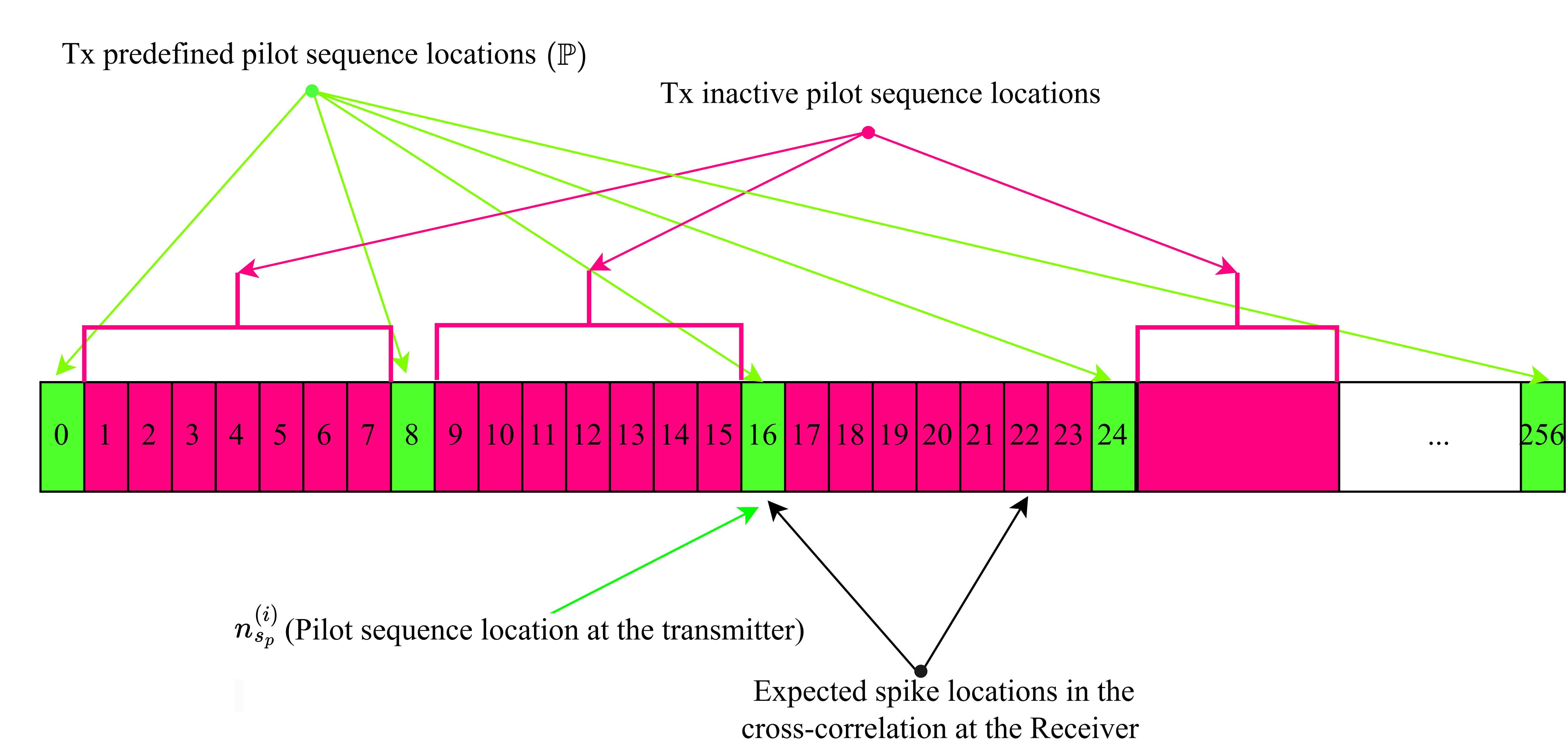} %
\caption{Example for PSP algorithm having pilot sequence located in position 8 over HF channel with 7 taps.}
\label{fig:44}
\end{figure*}
%%%%%%%%%%%%%%%%%%%%%%%%%%%%%

\section{Pilot sequence placement algorithm}

In this section, we introduce our PSP algorithm to incorporate additional information bits within the pilot sequence location of each frame. Not solely restricted to HF channels, the PSP algorithm at the transmitter can be applied to any channel. The frame structure employed in this paper is illustrated in Fig. \ref{fig:2}. 
{This frame structure considers} the detrimental effects of inter-block interference (IBI) which is discussed in detail in \cite{keykhosravi2023pilot}.
{As described in detail in the system model, the pilot sequence can be located at $n_{s_p}^{(i)}=\{0,1, ..., 2^{\lfloor {\log_2}{({{N_\text{s}}}+1)} \rfloor}-1\}$ in a given frame to carry ${\lfloor {\log_2}{({{N_\text{s}}}+1)} \rfloor}$ additional bits.}

Identifying the location of the pilot sequence, $\hat{n}_{s_p}^{(i)}$, at the receiver by the pilot sequence identification algorithm depends on the magnitude of the received signal for decision-making. This could be problematic for example when the first tap of the $(L_\text{c}+1)$-tap HF channel is in a deep fade. In a multipath channel, the received signal {consists of} a summation of different versions of the transmitted signal received from different paths of the channel. Thus, if the first path is in a deep fade, the first version of the signal becomes very weak at the receiver and the strong version of the signal coming from the strong tap becomes dominant in the receiver. Having said that, at the receiver, employing a correlation-based technique, the location of the peak that corresponds to the pilot sequence received through the strongest tap of the channel could be identified and not the location that corresponds to the first tap, $n_{s_p}^{(i)}$. To address this issue, in our proposed PSP algorithm, we restrict $n_{s_p}^{(i)}$ to some predefined locations. 
In other words, we do not allow $n_{s_p}^{(i)}$ to take all possible values in $\{0,1, ..., 2^{\lfloor {\log_2}{({{N_\text{s}}}+1)} \rfloor}-1\}$. 
{We define a set, denoted as $\mathbb{P}$, to represent the predefined locations $\{0, 1, \ldots, 2^{\lfloor \log_2({{N_\text{s}}}+1) \rfloor} - 1\}$ which are multiples of $2^{\lceil \log_2(L_c+1) \rceil}$. The cardinality of $\mathbb{P}$, $\text{Card}\{\mathbb{P}\}$, is equal to $2^{{\lfloor {\log_2}{({{N_\text{s}}}+1)} \rfloor}-\lceil{{\log_2}{(L_\text{c}+1)}}\rceil}$.
This represents the number of choices for $n_{s_p}^{(i)}$ at the transmitter. Consequently, the location of the pilot sequence could carry an additional $ N_\text{b} = {{\lfloor {\log_2}{({{N_\text{s}}}+1)} \rfloor}-\lceil{{\log_2}{(L_\text{c}+1)}}\rceil}$ bits.
{As $n_{s_p}^{(i)}$, a member of predefined locations set $\mathbb{P}$, is a multiple of $2^{\lceil \log_2(L_c+1) \rceil}$, can be represented with ${\lfloor {\log_2}{({{N_\text{s}}}+1)} \rfloor}$ bits where the ${\lceil \log_2(L_c+1) \rceil}$ least significant bits of its binary representation are zero.} Thus, $N_\text{b}$ additional data bits that are carried through the location of the pilot in a given frame can be mapped uniquely to a location in $\mathbb{P}$ where $N_\text{b}$ most significant bits (MSB) of the binary representation specify the location.
Consider a common HF channel having $L_\text{c}+1$ taps with two of them to be non-zero ($c_0\ \text{and}\ c_{L_\text{c}}$). For a given $n_{s_p}^{(i)}$ in a given frame $i$ at the transmitter belongs to the predefined locations set, $\mathbb{P}$, we expect to have spikes at the locations ${n}_{s_p}^{(i)}$ and $n_{s_p}^{(i)}+L_\text{c}$ in the square of the absolute value of the cross-correlation of the received signal with the pilot sequence, $|R_{\tilde{r}\tilde{p}} (\delta)|^2$ for $\delta = 0, ..., N - 1$, as defined in Section ~\RomanNumeralCaps{4} at the receiver. Thus, the set of expected pilot sequence locations at the receiver, denoted as $\hat{\mathbb{P}}$, consists of $\{{n}_{s_p}^{(i)},n_{s_p}^{(i)}+L_\text{c}\}$ for all ${n}_{s_p}^{(i)} \in \mathbb{P}$. Consequently, an estimated pilot sequence location at the receiver, $\hat{n}_{s_p}^{(i)}$, should belong to $\hat{\mathbb{P}}$ to be valid. As for each predefined pilot sequence location at the transmitter ${n}_{s_p}^{(i)} \in \mathbb{P}$, we have two expected pilot sequence locations the receiver, $\{{n}_{s_p}^{(i)},n_{s_p}^{(i)}+L_\text{c}\} \in \hat{\mathbb{P}}$, the cardinality of $\hat{\mathbb{P}}$, $\text{Card}\{\hat{\mathbb{P}}\}= 2 \times \text{Card}\{{\mathbb{P}}\}$. Thus, we have $\text{Card}\{\hat{\mathbb{P}}\} = 2\times 2^{{\lfloor {\log_2}{({{N_\text{s}}}+1)} \rfloor}-\lceil{{\log_2}{(L_\text{c}+1)}}\rceil}$.}

It is best to explain the idea behind the proposed PSP algorithm with the help of the following example. Consider a system with $N_\text{p} = 32$, ${{N_\text{s}}}=256$, and a channel with 7 taps ($ L_\text{c}=6$) with two non-zero taps ($c_0\ \text{and}\ c_{L_\text{c}}$) having a fixed delay spread with the first tap has lower strength. 
{For this example, we investigate two cases: 

Case 1: Consider the case that we allow $n_{s_p}^{(i)}$ to take all possible values in $\{0,1, ..., 2^{\lfloor {\log_2}{({{N_\text{s}}}+1)} \rfloor}-1\}$. As in this example ${\log_2}{({{N_\text{s}}}+1)}= 8.0056$ is not an integer number and thus $ {\lfloor{{\log_2}{({{N_\text{s}}}+1)}}\rfloor} = 8$ bits could be carried through the pilot sequence location. This is equivalent to $2^{\lfloor {\log_2}{({{N_\text{s}}}+1)} \rfloor}=256$ (which is less than ${{{N_\text{s}}}+1}=257$) choices for the pilot sequence to be located to construct the frame, and thus, the pilot sequence location $n_{s_p}^{(i)}=\{0,1, ..., 2^{\lfloor {\log_2}{({{N_\text{s}}}+1)} \rfloor}-1=255\}$. Let us for the first case, consider the pilot sequence’s location at the transmitter $n_{s_p}^{(i)}$ to be $10$, the receiver receives the summation of two versions of the signal corresponding to each non-zero channel tap. Fig. \ref{fig:1111} depicts the square of the absolute value of the cross-correlation of the received signal with the pilot sequence, i.e., $|R_{\tilde{r}\tilde{p}} (\delta)|^2$, $\delta = 0, ..., N - 1$, as defined in Section ~\RomanNumeralCaps{4}, under the presumption that the pilot sequence has perfect auto-correlation properties. As shown in Fig. \ref{fig:1111} there are two spikes in $|R_{\tilde{r}\tilde{p}} (\delta)|^2$, one at the location $n_{s_p}^{(i)}=10$ and the other at the location $n_{s_p}^{(i)}+L_\text{c}=16$ where the second is the strongest. Hence, the location of the pilot sequence $\hat{n}_{s_p}^{(i)} $ will be incorrectly identified to be $n_{s_p}^{(i)}+L_\text{c}=16$ if the PSP algorithm is not designed carefully, i.e. $n_{s_p}^{(i)}$ does not belong to $\mathbb{P}$. As mentioned before, to address this issue, we do not allow $n_{s_p}^{(i)}$ to take all possible values in $\{0,1, ..., 255\}$. The predefined locations should belong to $\mathbb{P}$. In this example, $\mathbb{P}$ consisted of integer values in  $\{0,1, ..., 2^{\lfloor {\log_2}{({{N_\text{s}}}+1)} \rfloor}-1=255\}$ which are multiples of $2^{\lceil{{\log_2}{(L_\text{c}+1)}}\rceil} = 2^{\lceil{{\log_2}{7}}\rceil}=8$. Thus, in this example, $n_{s_p}^{(i)}$ should have been chosen to  belong to the set $\mathbb{P} =\{0, 2^{\lceil{{\log_2}{(L_\text{c}+1)}}\rceil}, 2\times2^{\lceil{{\log_2}{(L_\text{c}+1)}}\rceil}, 3\times2^{\lceil{{\log_2}{(L_\text{c}+1)}}\rceil}, ..., 31\times 2^{\lceil{{\log_2}{(L_\text{c}+1)}}\rceil}\}= \{0, 8, 2\times8, 3\times8, ..., 248\}$. Having said that, as $n_p^{(i)}=10$ does not belong to $\mathbb{P}$, thus we can not set $n_p^{(i)}$ to be $10$ at the transmitter.}

{Fig \ref{fig:44} further illustrates an example of pilot locations allowed by PSP algorithm. Predefined pilot sequence locations, where the transmitter can place the pilot sequence within a frame, are represented in green. Blocked locations, which are unavailable for the transmitter to use for pilot sequence placement, are depicted in pink. For a given $n_{s_p}^{(i)}$ in a given frame $i$ at the transmitter, we expect to have spikes at the locations ${n}_{s_p}^{(i)}$ and $n_{s_p}^{(i)}+L_\text{c}$ in $|R_{\tilde{r}\tilde{p}} (\delta)|^2$ at the receiver. Thus the estimated pilot sequence location at the receiver $\hat{n}_{s_p}^{(i)}$ should belong to the set $\hat{\mathbb{P}} =\{\{0,6\}, \{8,14\}, \{16,22\}, \{24,30\}, ..., \{248,256\}\}$.} 

{Case 2: Let us consider $n_{s_p}^{(i)} = 16$ belongs to $\mathbb{P}$ at the transmitter.} Now having the described channel, at the receiver we expect two spikes $n_{s_p}^{(i)}=16$ and $n_{s_p}^{(i)}+L_\text{c}=22$  (as shown in Fig. \ref{fig:44}) with the second spike to be stronger. Based on $|R_{\tilde{r}\tilde{p}} (\delta)|^2$, the location of the pilot sequence location is identified to be $n_{s_p}^{(i)}+L_\text{c}=22$ but we know that the location of the pilot sequence $n_{s_p}^{(i)}$ is a multiple of 8 and thus $n_{s_p}^{(i)}$ module 8 should be zero. Thus, as 22 belongs to the expected locations $\{16,22\}$ we can identify $\hat{n}_{s_p}^{(i)}=16$. In other words, 22 module 8 is 6, and by subtracting 6 from 22, we can correctly identify $\hat{n}_{s_p}^{(i)}=16$. In this example, there are $2^{{\lfloor {\log_2}{256+1} \rfloor}-\lceil{{\log_2}{(6+1)}}\rceil}=2^{8-3}=32$ choices for $n_{s_p}^{(i)}$ at the transmitter and thus the location of pilot sequence could carry $ N_\text{b} = {{\lfloor {\log_2}{256+1} \rfloor}-\lceil{{\log_2}{(6+1)}}\rceil}= 8-3 =5$ additional bits in each frame.

The SE measured in bits/sec/Hz in a Nyquist system, FTN signaling system and our proposed IM-based channel estimation for FTN signaling system is denoted respectively as $\gamma_{\text{Nyq}}$, $\gamma_{\text{FTN}}$, and $\gamma_{\text{IM-FTN}}$ and is defined as 
\begin{IEEEeqnarray}{rcl}\label{equationSE}
\gamma_{\text{Nyq}} &{}={}&  \frac{{{N_\text{s}}}}{({{N_\text{s}}}+N_\text{p})} \cdot\frac{{\log_2}M}{(1+\beta)} {{R_\text{c}}},\\
\gamma_{\text{FTN}} &{}={}&  \frac{{{N_\text{s}}}}{({{N_\text{s}}}+N_\text{p})} \cdot\frac{{\log_2}M}{\tau (1+\beta)}R_\text{c},\\
\gamma_{\text{IM-FTN}} &{}={}&  {\frac{N_\text{s}\cdot R_\text{c}\cdot{{\log_2}M}+{N_\text{b}}}{(N_\text{s}+N_\text{p})} \cdot\frac{1}{\tau (1+\beta)}},
\end{IEEEeqnarray}
where $\beta$ is the roll-off factor of the RRC pulse shaping filter and {$R_\text{c}$ is the code rate}. It should be noted that in our proposed IM-based channel estimation for the FTN signaling system, $N_\text{b}$ additional bits are transmitted in a given frame. The improvement percentage of the SE of the proposed IM-based channel estimation for the FTN signaling over Nyquist signaling and over FTN signaling is defined, respectively, as

\begin{IEEEeqnarray}{rcl}\label{equationImprov}
g_{\text{IM-FTN,Nyq}} &{}={}&  \frac{(\gamma_{\text{IM-FTN}}-\gamma_{\text{Nyq}})}{\gamma_{\text{Nyq}}}  \nonumber\\
&{}={}&{{\frac{{({N_\text{s}}\cdot R_\text{c}\cdot{{\log_2}M})}(\frac{1}{\tau}-1)+\frac{N_\text{b}}{\tau}}{N_\text{s}\cdot R_\text{c} \cdot {{\log_2}M}}}} \times 100 \%, \nonumber\\
g_{\text{IM-FTN,FTN}} &{}={}&  \frac{(\gamma_{\text{IM-FTN}}-\gamma_{\text{FTN}})}{\gamma_{\text{FTN}}} = \frac{N_\text{b}}{{{N_\text{s}\cdot R_\text{c}}} \cdot{{\log_2}M}} \times 100\% . \nonumber\\
\end{IEEEeqnarray}

Table \ref{tab:table_1} shows the spectral efficiencies $\gamma_{\text{Nyq}}$, $\gamma_{\text{FTN}}$, and $\gamma_{\text{IM-FTN}}$ and the percentage of SE improvements for our proposed IM-based channel estimation for the FTN signaling compared to Nyquist signaling $g_\text{IM-FTN,Nyq}$ and the FTN signaling $g_\text{IM-FTN,FTN}$ for practical values of ${{N_\text{s}}}$ and $N_\text{p}$ employed in HF systems for $\beta=0.35$ and $\tau=0.72$, {when considering the $R_\text{c}=3/4$}. As shown in Table \ref{tab:table_1}, there could be more than 40\% improvement in SE when utilizing the proposed IM-based channel estimation for the FTN signaling compared to the traditional Nyquist signaling for some typical values ${{N_\text{s}}}$ and $N_\text{p}$ for HF systems. As well as that Table \ref{tab:table_1} shows that there could be between approximately {1.3}\% and {5.55}\% improvement in SE when utilizing the proposed IM-based channel estimation for FTN signaling compared to the FTN signaling with the conventional pilot sequences (pilot sequence location being fixed in the frames) for some typical values ${{N_\text{s}}}$ and $N_\text{p}$ for HF systems.

\begin{table*}[t]
  \centering
    \caption{SE gains employing the proposed IM-based channel estimation for FTN signaling system for {$\beta=0.35$, $\tau=0.72$, and $R_\text{c}=3/4$}}.
  \begin{tabular}{|c|c|c|c|c|c|c|c|}
    \hline
    $N_\text{p} \: (\text{Symbols})$ & ${{N_\text{s}}} \: (\text{Symbols})$ & $M$ & $\gamma_{\text{Nyq}} \: ({\text{bits/sec/Hz}})$ &$\gamma_{\text{FTN}}$ \:({\text{bits/sec/Hz}})& $\gamma_{\text{IM-FTN}}$ \:({\text{bits/sec/Hz}})& $g_\text{IM-FTN,Nyq}$ & $g_\text{IM-FTN,FTN}$\\
    \hline
    48 & 48 & 2 & {0.2778}& {0.3858} & {0.4072}  & {46.60}\% & {5.55}\% \\
    \hline
    32 & 96 & 2 & {0.4167} & {0.5787} & {0.6028} & {44.67}\% & {4.16}\% \\
    \hline
    32 & 256 & 2 & {0.4938} &{0.6859} & {0.7037} & {42.50}\% & {2.60}\% \\
    \hline
    32 & 256 & 4 & {0.9877} &{1.3717} & {1.3896} & {40.70}\% & {1.30}\% \\
    \hline
  \end{tabular}
  \label{tab:table_1}
\end{table*}

\subsection{Pilot sequence Placement Algorithm}
The proposed pilot sequence placement algorithm is summarized as:

\begin{algorithm}
  \caption{Pilot sequence Placement Algorithm}
  \textbf{Input}:   ${{N_\text{s}}}$ and $L_\text{c}$\\
  1) Construct the set of predefined pilot sequence locations allowed at the transmitter, ${\mathbb{P}}$, consisting of integer multiples of $2^{\lceil \log_2(L_\text{c}+1) \rceil}$ belonging to the set $\{0,1, ..., 2^{\lfloor {\log_2}{({{N_\text{s}}}+1)} \rfloor}-1\}$.\\
  2) Calculate the number of bits that can be carried through the pilot sequence location $ N_\text{b} = {{\lfloor {\log_2}{({{N_\text{s}}}+1)} \rfloor}-\lceil{{\log_2}{(L_\text{c}+1)}}\rceil}$.   \\
  3) Map the $N_\text{b}$ additional data bits to a predefined location in ${\mathbb{P}}$, $n_{s_p}^{(i)}$, where the $N_\text{b}$ MSB of the binary representation specify the location.\\
  4) Place the pilot sequence of length $N_\text{p}$ in the location $n_{s_p}^{(i)}$ of the $i$th frame of length $N$. \\
  \textbf{Output}:  $n_{s_p}^{(i)}$
\end{algorithm}

\section{Pilot Sequence Location Identification}
In this section, we discuss the proposed low-complexity pilot sequence identification algorithm to identify the location of the pilot sequence in each frame at the receiver. As mentioned earlier, this paper focuses on the HF channel model recommended by the International Telecommunications Union Radio Communication Sector (ITU-R). The ITU-R has %furnished 
{introduced} a comprehensive set of recommended HF channel models widely utilized for simulating the performance of HF systems. ITU-R F.520 \cite{ITU} %provides a recommendation encompassing 
{recommends} three distinct test channels, namely the ITU-R good, ITU-R moderate, and ITU-R poor channel, which serve to classify the channel conditions into varying levels based on their individual characteristics. Although these channels are still commonly utilized for assessing the performance of HF systems, it is important to note that the corresponding recommendation has become outdated and has been %superseded 
{replaced} by ITU-R F.1487 \cite{ITUF}. ITU-R F.1487 introduces a comprehensive set of 10 test channels, which {include} various latitude regions and levels of ionospheric disturbance. Notably, each of the three test channels specified in F.520 aligns with at least one of the test channels defined in F.1487. 

The ITU-R F.520 \cite{ITU} and ITU-R F.1487 \cite{otnes2002improved} recommendations describe all test channels as tapped delay line models comprising only two non-zero taps. These taps exhibit independent fading, undergoing a Rayleigh probability density function. Additionally, both paths experience equal average power with the same Doppler frequency. In this paper, the ITU-R poor channel is utilized for simulation purposes to represent worst case scenario on channel estimation performance. The channel condition labeled as poor channel is in accordance with the ITU-R F.1487 specifications for the mid-latitude-disturbed channel. It comprises two non-zero independent paths having a fixed delay of 2 ms and a frequency Doppler of 1 Hz. The channel is assumed to have a delay spread of 2.1 ms, allowing for precisely 5 symbol intervals in the Nyquist case. Considering a symbol rate of 2400 symbols per second, this channel can be represented as a Watterson model with $6$ taps.

To {determine} the pilot sequence's location within a given receiver frame, we employ a cross-correlation technique. This approach serves two primary purposes: firstly, it {avoids} the excessive complexity associated with the maximum likelihood (ML) technique required to identify the pilot sequence location, and secondly, it {utilizes} the intrinsic {characteristics} of both the HF channel and the favorable local auto-correlation of the optimal pilot sequence designed to minimize the mean square error (MSE) of channel estimation. The proposed pilot sequence identification algorithm involves correlating the output of the whitening filter in the receiver, $\tilde{r}_k$, with the known pilot sequence, which has been subject to FTN-induced ISI and the whitening filter. 
In this study, we utilize an optimized pilot sequence that minimizes the MSE of channel estimation for FTN signaling over a doubly-selective channel, as proposed in \cite{keykhosravi2023pilot}. 
The pilot sequence affected by FTN-induced ISI and a whitening filter can be characterized as 
\begin{IEEEeqnarray}{rcl}\label{equation31}
\tilde{p}_{k} &{}={}& \sum \limits_{j = 0}^{N_p-1}  {p_j  v_{k-j} },   \quad   k = 0, ... , N_p-1.
\end{IEEEeqnarray}

%%%%%%%%%%%%%%%%%%%%%%%%
\begin{figure}[t]
\centering
\includegraphics[width=8.5cm,height=6cm]{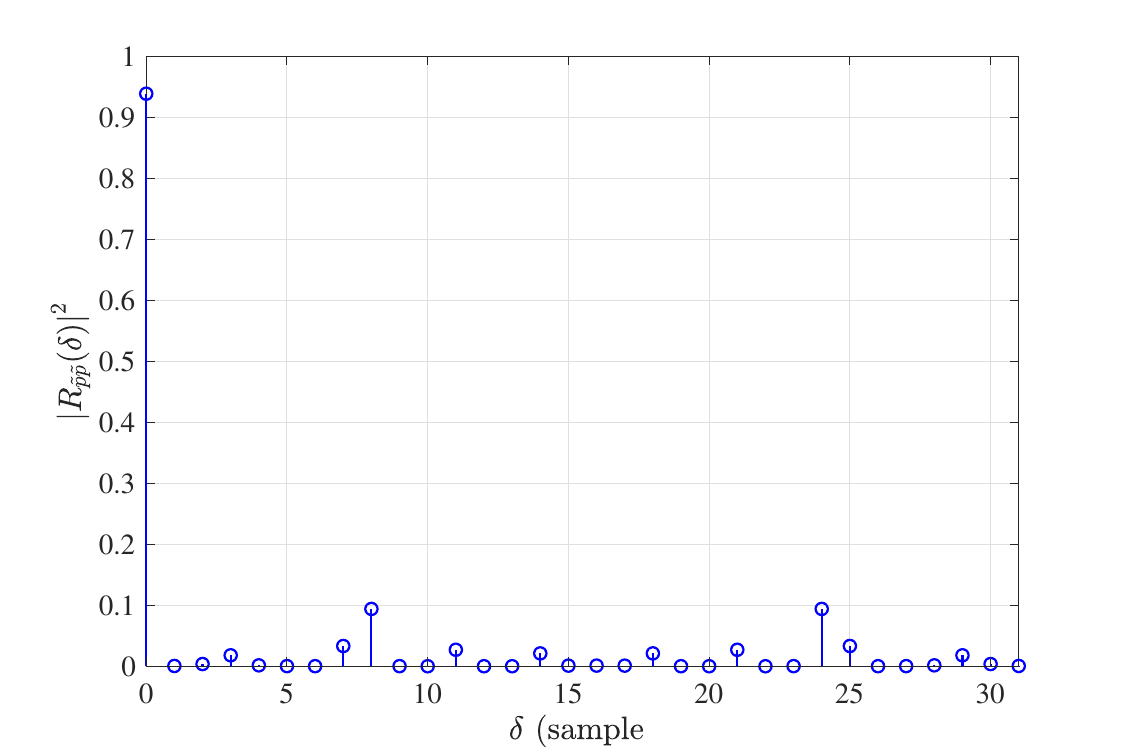}
\caption{The square of the absolute value of the autocorrelation of the pilot sequence of length $N_\text{p}=32$ affected by FTN-induced ISI and whitening filter.}
\label{fig:3}
\end{figure}

\noindent{The value of the autocorrelation of the designed pilot sequence affected by FTN-induced ISI and whitening filter indicated in (\ref{equation31}) can be expressed as}
\begin{IEEEeqnarray}{rcl}\label{equation32}
R_{\tilde{p}\tilde{p}} (\delta) &{}={}&  \sum \limits_{k = 0}^{N_p-1} { \tilde{p}_{k} \tilde{p^\ast}_{\delta+k}},  \quad  \delta = 0, ... , N_p-1.
\end{IEEEeqnarray}
Fig. \ref{fig:3} presents the square of the absolute value of the autocorrelation function ${|R_{\tilde{p}\tilde{p}} (\delta)|}^2$ of the pilot sequence designed for channel estimation of FTN signaling with $\tau = 0.84$ over a discretized doubly-selective HF channel. The choice of $\tau = 0.84$ ensures that the number of taps in the tapped delay line channel model is an integer. 
As can be seen in Fig. \ref{fig:3}, the optimal pilot sequence has desirable local auto-correlation characteristics that assist in the pilot sequence identification. The local auto-correlation characteristics $\psi (\delta)$ at a given location $\delta$, $\delta = 0, ... , N_p-1.$ is defined as the ratio of ${|R_{\tilde{p}\tilde{p}} (\delta)|}^2$ and the normalized sum of ${|R_{\tilde{p}\tilde{p}} (j)|}^2$ for adjacent locations $j$ within a given radius $r_0$, and the value of $r_0$ should be chosen to be less than $L_\text{c}$ to avoid taking into account the contribution of the second channel tap. Hence, the local auto-correlation characteristics $\psi (\delta)$ is formally defined as
\begin{IEEEeqnarray}{rcl}\label{equation32_2}
	\psi (\delta) &{}={}&  
	\frac{{|R_{\tilde{p}\tilde{p}}(\delta)|}^2}{\lambda (\delta)}, \quad \delta = 0, ... , N_p-1,
\end{IEEEeqnarray}
where 
\begin{IEEEeqnarray}{rcl}\label{equation32_1}
	\lambda (\delta) &{}={}& \frac{{\sum \limits_{j = \max(\delta-r_0,\delta_{\text{min}})}^{\min(\delta+r_0,\delta_{\text{max}})}{|R_{\tilde{p}\tilde{p}}(j)|}^2-{|R_{\tilde{p}\tilde{p}}(\delta)|}^2}}{{ \min(\delta+r_0,\delta_{\text{max}})-\max(\delta-r_0,\delta_{\text{min}})} },    
\end{IEEEeqnarray}
where $\delta_{\text{min}}=0$ and $\delta_{\text{max}}={{N_\text{s}}}$ are the lower and upper limits of $\delta$ in ${|R_{\tilde{p}\tilde{p}} (\delta)|}^2$, respectively.

We utilize the characteristics of $|R_{\tilde{p}\tilde{p}}(\delta)|^2$ to define a measure for determining the pilot sequence's location within a given frame based on the square of the absolute of the cross-correlation between the received signal $\tilde{r}_{k}$ and the pilot sequence affected by FTN-induced ISI and the whitening filter, denoted as $\tilde{p}_{k}$. Considering the received signal for a given frame $i$ in  (\ref{equation7}) as a sequence of $\tilde{r}_{k}, k = 0,..., {{N_\text{s}}}-1$, the cross-correlation of $\tilde{r}_{k}$ and $\tilde{p}_{k}$ can be written as
\begin{IEEEeqnarray}{rcl}\label{equation34}
R_{\tilde{r}\tilde{p}} (\delta) &{}={}&  \sum \limits_{m = 0}^{N_p-1} { \tilde{r}_{\delta+m} \tilde{p}_{m}^\ast},  \quad  \delta = 0, ... , N_d-1, \nonumber\\
&{}={}& \sum \limits_{m = 0}^{N_p-1} \sum \limits_{i=1}^{N_i}  \left(\tilde{p}_{m}^\ast \mkern-12mu \sum \limits_{n_p^{(i)} \in \Omega_\text{p}^{(i)}} \mkern-18mu {s_{n_p^{(i)}} \sum\limits_{l=0}^{L_\text{c}} c_{\delta+m,l} v_{\delta+m-l-n_p^{(i)}} }\right)  \nonumber \\
& &{}+  \sum \limits_{m = 0}^{N_p-1} \sum \limits_{i=1}^{N_i} 
  \left(\tilde{p}_{m}^\ast \mkern-18mu \sum \limits_{n_d^{(i)} \in \Omega_\text{d}^{(i)}} \mkern-18mu {s_{n_d^{(i)}}  \sum\limits_{l=0}^{L_\text{c}} c_{\delta+m,l} v_{\delta+m-l-n_d^{(i)} }}\mkern-8mu\right) \nonumber \\
&& {}+{} \sum \limits_{m = 0}^{{{N_\text{s}}}-1} w_{\delta+m}  \tilde{p}_{m}^\ast. \IEEEeqnarraynumspace
\end{IEEEeqnarray}

To simplify the notation, let us consider the transmission of one frame and drop the frame index $i$. Having the cross-correlation of $\tilde{r}_{k}$ and $\tilde{p}_{k}$ in (\ref{equation34}), we could write the square of the absolute value of the $|R_{\tilde{r}\tilde{p}} (\delta)|^2$ over a general frequency selective time-invariant channel as
\begin{IEEEeqnarray}{rcl}\label{equation34_2}
\small{|R_{\tilde{r}\tilde{p}} (\delta)|}^2 &=&  \nonumber \\
&&\mkern-12mu\small\begin{cases}
\small{|c_l|}^2 {|R_{\tilde{p}\tilde{p}} (0)|}^2 \!+\! \zeta(\delta),   \delta=n_{s_p}+l, l=\{0, 1, ..., L_\text{c}\},\vspace{12pt}\mkern-2mu\\
\small\zeta(\delta),\qquad \qquad \qquad \ \ \text{otherwise.}
\end{cases} \nonumber \\
\IEEEeqnarraynumspace
\end{IEEEeqnarray}
The proof of \eqref{equation34_2} is provided in the appendix. 
Hence, for the received symbols corresponding to the pilot sequence, the square absolute value of the cross-correlation of received symbols and pilot sequence, i.e., ${|R_{\tilde{r}\tilde{p}} (\delta)|}^2$, for 
$\delta=n_{s_p}+l, \: l=\{0, 1, ..., L_\text{c}\}$ is the ${|R_{\tilde{p}\tilde{p}} (0)|}^2$ weighted by the gain of corresponding channel path. In other words, the value of ${|R_{\tilde{r}\tilde{p}} (\delta)|}^2$ at $\delta=n_{s_p}$ is equal to ${|R_{\tilde{p}\tilde{p}} (0)|}^2$ scaled by ${|c_0|}$ and an interfere term $\zeta(\delta)$.

For the HF poor channel having $c_0$ and $c_{L_\text{c}}$ to be the two non-zero taps, we rewrite \eqref{equation34_2} as
\begin{IEEEeqnarray}{rcl}\label{equation34_3}
{|R_{\tilde{r}\tilde{p}} (\delta)|}^2 &{}={}&  
\begin{cases}
{|c_0|}^2 {|R_{\tilde{p}\tilde{p}} (0)|}^2 + \zeta(\delta), & \qquad \delta=n_{s_p},\\
{|c_{L_\text{c}}|}^2 {|R_{\tilde{p}\tilde{p}} (0)|}^2 + \zeta(\delta), & \qquad \delta=n_{s_p}+c_{L_\text{c}},\\
\zeta(\delta), & \qquad\text{otherwise}.
\end{cases} \nonumber \\
\IEEEeqnarraynumspace
\end{IEEEeqnarray} 

To {explain} the proposed algorithm, we initially describe the proposed algorithm through different time-invariant HF channel realizations. 
It is important to highlight that a potential approach for determining the pilot sequence location involves {considering} only the maximum value of $|R_{\tilde{r}\tilde{p}} (\delta)|^2$. However, it fails to {utilize the inherent characteristics} of the HF channel delay spread and does not lead to improved performance. 
In order to enhance the accuracy of pilot sequence location identification, we propose an algorithm that {includes a measure that accounts} for the {unique} characteristics of the HF channel and the favorable square of the absolute value of the auto-correlation of the optimal pilot sequence as discussed in the next subsection.

\begin{figure}[t]
\centering
\includegraphics[width=8.5cm,height=6cm]{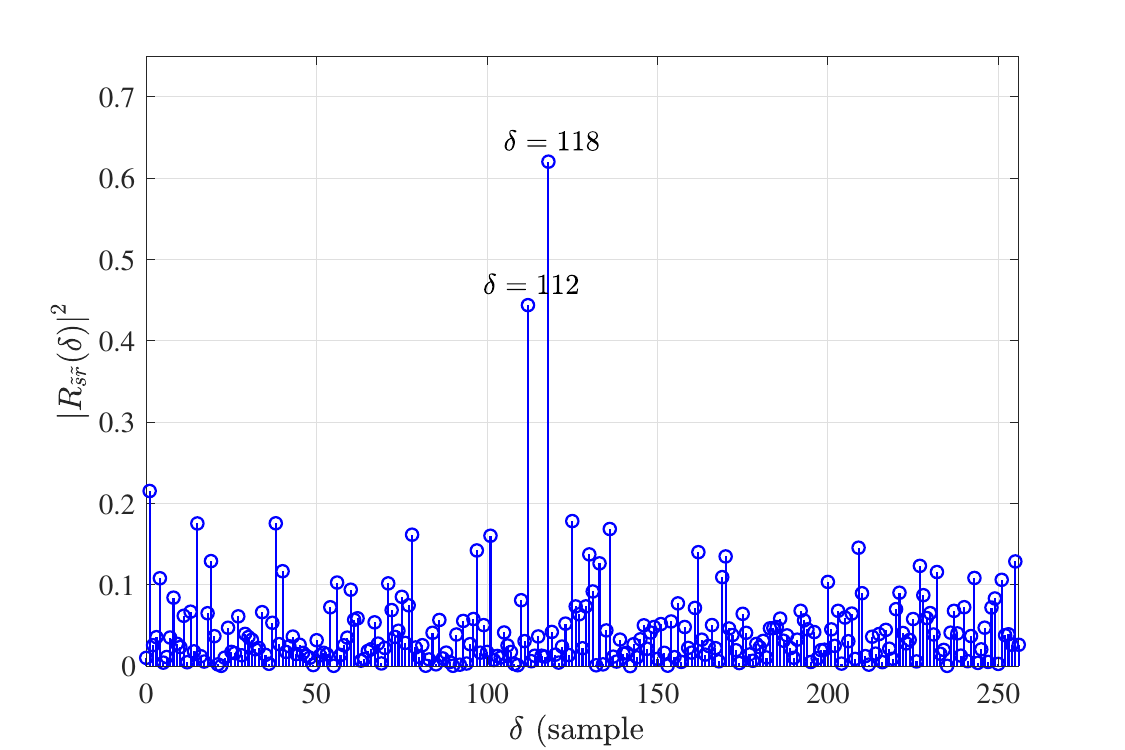}
\caption{$|R_{\tilde{r}\tilde{p}} (\delta)|^2$ with $N_\text{p}=32$, ${{N_\text{s}}}=256$, $\textbf{b}=[0,1,1,1,0]$, ${|\textbf{c}|}^2=[0.30,0,0,0,0,0,0.699]^\text{T}$, and possible pilot sequence location from 0 to 256.}

\label{fig:6}
\end{figure}
\subsection{Pilot Sequence Location Identification for Time-Invariant HF Channel}

In this subsection, we introduce the proposed algorithm for pilot sequence location identification in HF channels. To better illustrate the algorithm, we provide examples using some selected realizations of the time-invariant HF channels.

Fig. \ref{fig:6} illustrates $|R_{\tilde{r}\tilde{p}} (\delta)|^2$ for the scenario where the transmit frame has $N_\text{p}=32$ and ${{N_\text{s}}}=256$.
 When considering a typical symbol rate of 2400 in HF systems, the delay spread of 2.1 ms in the FTN signaling system, with a value of $\tau = 0.84$, can be equivalently represented by a discretized channel with 7 taps. 
$ N_\text{b} = {{\lfloor {\log_2}{({{N_\text{s}}}+1)} \rfloor}-\lceil{{\log_2}{(L_\text{c}+1)}}\rceil} = {{\lfloor {\log_2}{(256+1)} \rfloor}-\lceil{{\log_2}{(6+1)}}\rceil}=5$. Thus, as explained in detail in Section ~\RomanNumeralCaps{3}, in this example, $n_{s_p}^{(i)}$ is predefined to be multiples of $2^{\lceil{{\log_2}{(L_\text{c}+1)}}\rceil}$ belong to $\{0,1, ..., 2^{\lfloor {\log_2}{({{N_\text{s}}}+1)} \rfloor}-1\}$. Thus, $n_{s_p}^{(i)}$ belongs to the set $\mathbb{P} = \{0, 2^{\lceil{{\log_2}{(L_\text{c}+1)}}\rceil}, 2\times2^{\lceil{{\log_2}{(L_\text{c}+1)}}\rceil}, 3\times2^{\lceil{{\log_2}{(L_\text{c}+1)}}\rceil}, ..., 31\times 2^{\lceil{{\log_2}{(L_\text{c}+1)}}\rceil}\}= \{0, 8, 2\times8, 3\times8, ..., 248\}$. As the length of the predefined locations to choose $n_{s_p}^{(i)}$ from is $\text{Card}\{\mathbb{P}\} = 32$ (designed to be $2^{N_\text{b}}$), {we map each bit sequence of length $N_\text{b}=5$ to one predefined location in $\mathbb{P}$ and place the pilot sequence in that location to construct the frame. At the transmitter, we consider the example when $\textbf{b}=[0,1,1,1,0]$ and map it to a decimal equivalent of 112 (a multiple of 8 whose $N_\text{b}=5$ MSB of the binary representation is equal to $\textbf{b}$), belongs to the predefined pilot sequence set of $\mathbb{P}$.}
Consider the channel realization given by ${|\textbf{c}|}^2=[0.30,0,0,0,0,0,0.699]^\text{T}$, with the 7th channel tap being the strongest. For a given $n_{s_p}^{(i)}$ in a given frame $i$ at the transmitter, we expect to have spikes at the locations $\hat{n}_{s_p}^{(i)}$ and $n_{s_p}^{(i)}+L_\text{c}$ in the square of the absolute value of the cross-correlation at the receiver. Thus $\hat{n}_{s_p}^{(i)}$ belongs to $\hat{\mathbb{P}} = \{\{0,6\}, \{8,14\}, ..., \{112,118\}, ..., \{248,256\}\}$. 
Having $n_{s_p}^{(i)}=112$, at the receiver we expect two spikes $n_{s_p}^{(i)}=112$ and $n_{s_p}^{(i)}+L_\text{c}=118$ with the second spike to be stronger. The peak location of ${|R_{\tilde{p}\tilde{p}} (\delta)|}^2$ in Fig. \ref{fig:6} is found to be 118. We know that the location of the pilot sequence $n_{s_p}^{(i)}$ is a multiple of 8 and thus $n_{s_p}^{(i)}$ module 8 should be zero. Thus, as 118 belongs to the expected locations $\{112,118\}$ we can identify $\hat{n}_{s_p}^{(i)}=112$. In other words, 118 module 8 is 6, and by subtracting 6 from 118, we can correctly identify $\hat{n}_{s_p}^{(i)}=112$.

%%%%%%%%%%%%%%%%%%%%%%%%%%%
\begin{figure}[t]
	\centering
	\includegraphics[width=8.5cm,height=6cm]{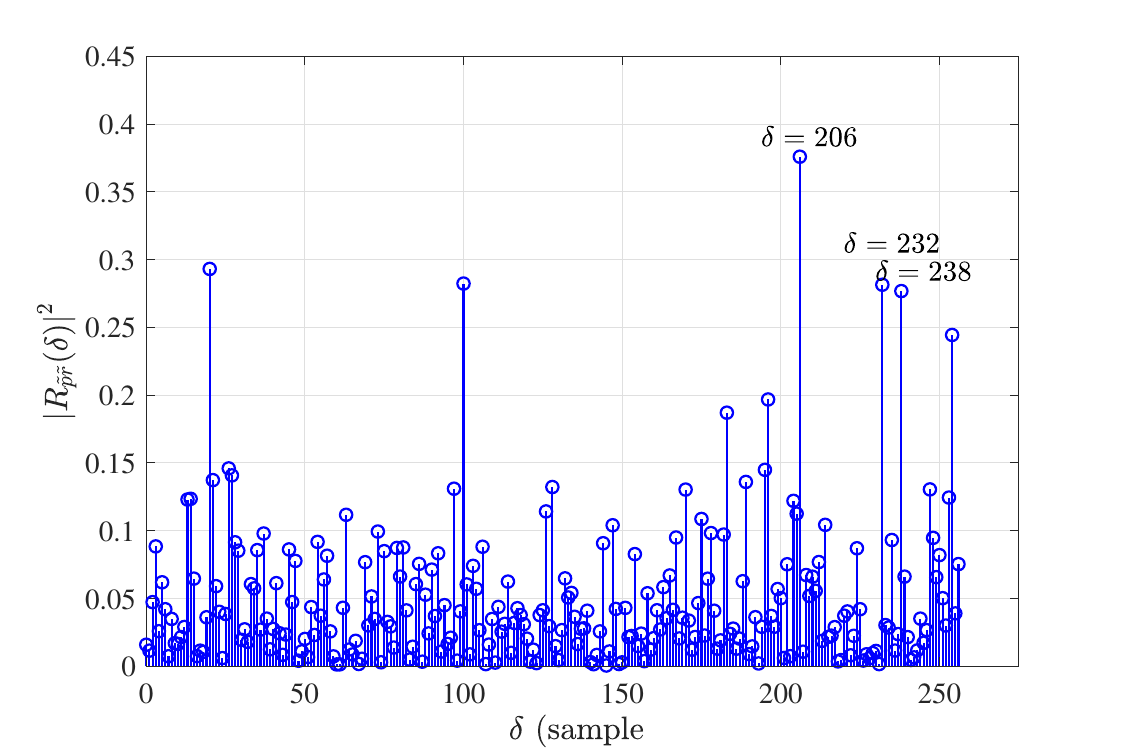}
	\caption{$|R_{\tilde{r}\tilde{p}} (\delta)|^2$ with $N_\text{p}=32$ and ${{N_\text{s}}}=256$, $\textbf{b}=[1,1,1,0,1]^\text{T}$, ${|\textbf{c}|}^2=[0.655,0,0,0,0,0,0.345]^\text{T}$, and possible pilot sequence location from 0 to 256.}
	\label{fig:5}
\end{figure}
%%%%%%%%%%%%%%%%%%%%%%%%%%%%%%%%%%%%%%%%%%%%%%%%%
Fig. \ref{fig:5} depicts the value of $|R_{\tilde{r}\tilde{p}} (\delta)|^2$ for the scenario where the transmit frame has $N_\text{p}=32$ and ${{N_\text{s}}}=256$. {At the transmitter, we consider the example when $\textbf{b}=[1,1,1,0,1]^\text{T}$ and map to its decimal worth of location 232 (a multiple of 8 whose $N_\text{b}=5$ MSB of the representation is equal to $\textbf{b}$), belongs to the predefined pilot sequence set of $\mathbb{P}$.} 
Consider the channel realization given by ${|\textbf{c}|}^2=[0.655,0,0,0,0,0,0.345]^\text{T}$, with the first channel tap being the strongest.
Similar to the previous example, the set of all expected pilot sequence locations at the receiver is $\hat{\mathbb{P}} = \{\{0,6\}, \{8,14\}, ..., \{112,118\}, ..., \{248,256\}\}$. 
The peak location of ${|R_{\tilde{p}\tilde{p}} (\delta)|}^2$ in Fig. \ref{fig:5} is found to be 206. We consider all expected locations where the ${|R_{\tilde{r}\tilde{p}} (\delta)|}^2$ surpasses a certain threshold $r_{\text{th}_1}$ which is equal to a constant value of $c_1$ multiply by the peak location of ${|R_{\tilde{p}\tilde{p}} (\delta)|}^2$.  
Thus, the algorithm identifies locations 206, 232, and 238 from the set of expected pilot sequence locations of $\hat{n}_{s_p}^{(i)}$ with a higher $|R_{\tilde{p}\tilde{p}} (\delta)|^2$ value than $r_{\text{th}_1}$. 
As 232 and 238 belong to $\{232,238\}$ in the set for $\hat{n}_{s_p}^{(i)}$ and corresponds to ${n}_{s_p}^{(i)}=232$, we need to consider only one of them, let us consider 232 between these two.
Subsequently, the proposed measure is calculated for locations 206 and 232 to select the location with a better local characteristic of the ${|R_{\tilde{r}\tilde{p}} (\delta)|}^2$. More specifically, we introduce a measure that {considers} the unique characteristics of the HF channel, characterized by only two non-zero taps separated by a specific delay spread. 
For any given location, the local characteristic of the square of the absolute value of the autocorrelation can be quantified as defined in (\ref{equation32_2}) by comparing the ratio of the ${|R_{\tilde{p}\tilde{p}} (\delta)|}^2$ of that location to the normalized sum of ${|R_{\tilde{p}\tilde{p}} (j)|}^2$ for $j$ adjacent locations within a radius. To accomplish this, we establish a radius, denoted as $r_0$ centered around potential locations identified by the algorithm. Based on the specifications of the system, we choose $r_0$ to be less than the length of the channel ($L_\text{c}+1$). This precaution is taken due to the expectation that the second non-zero tap will fall within a radius equal to the length of the channel ($L_\text{c}+1$).

In order to exploit the unique characteristics of the HF channel, we define the measure by considering the local characteristic of the ${|R_{\tilde{r}\tilde{p}} (\delta)|}^2$ for both non-zero taps jointly. To end this, we determine $\delta_{c_0}$ and $\delta_{c_{L_c}}$ as the corresponding locations for the two spikes in ${|R_{\tilde{r}\tilde{p}} (\delta)|}^2$ corresponding to two versions of the signal received from the two non-zero taps of the channel. As ${|R_{\tilde{r}\tilde{p}} (\delta)|}^2$ for $\delta_{c_0}$ and $\delta_{c_{L_c}}$ have high values, we add these two values in the numerator of the proposed measure $\mu (\delta_{c_0},\delta_{c_{L_c}})$.  
The normalized sum of ${|R_{\tilde{r}\tilde{p}} (\delta)|}^2$ for the adjacent locations within a radius $r_0$ around location $\delta_{c_0}$ and $\delta_{c_{L_c}}$ is defined as $\lambda({\delta_{c_0}})$ and $\lambda({\delta_{c_{L_c}}})$, respectively, as defined in (\ref{equation32_1}). Hence, the proposed measure $\mu (\delta_{c_0},\delta_{c_{L_c}})$ is defined as follows
\begin{IEEEeqnarray}{rcl}\label{equation36}
\mu (\delta_{c_0},\delta_{c_{L_c}}) = \frac{{|R_{\tilde{p}\tilde{r}}(\delta_{c_0})|}^2+{|R_{\tilde{p}\tilde{r}}(\delta_{c_{L_c}})|}^2}{\lambda({\delta_{c_0}})+\lambda({\delta_{c_{L_c}}})}.
\end{IEEEeqnarray}
For the given realization illustrated in Fig. \ref{fig:5}, upon examining location 206, we determine the congruence modulo $8$ to be 6. Consequently, we subtract the congruence modulo value from the location obtained through the measure, resulting in $\delta_{c_0}$ and $\delta_{c_{L_c}}$ being identified as 200 and 206, respectively. Following this, we proceed to calculate the proposed measure. Notably, since location 206 corresponds to the peak of ${|R_{\tilde{p}\tilde{r}} (\delta)|}^2$, we derive $r_{\text{th}_2}$ based on this particular observation as $r_{\text{th}_2} = c_2 \cdot \mu (200,206)$. Where $c_2$ is a constant value. Afterward, for location 232, within the same realization depicted in Fig. \ref{fig:5} we ascertain $\delta_{c_0}$ and $\delta_{c_{L_c}}$ to be 232 and 238, respectively, and subsequently compute the measure in \eqref{equation36}. The measure calculated for location 232 exceeds the established threshold. 
Consequently, the proposed algorithm successfully identifies 232 as the pilot sequence location within the current frame. It is important to note that relying {only} on the identification of the maximum of ${|R_{\tilde{r}\tilde{p}} (\delta)|}^2$ would lead to an incorrect determination of the pilot sequence location for this particular channel realization. However, through the utilization of the proposed algorithm, we successfully {address} this issue and achieve precise identification of the pilot sequence location.

Following the identification of the pilot sequence's location, the receiver proceeds to utilize the LSSE-based approach for estimating the complex channel coefficients of FTN signaling at the designated pilot sequence locations within the frame. Let us consider the received samples in (\ref{equation7}) corresponding to transmitted pilot sequence symbols for a given frame $i$. 
As explored in \cite{keykhosravi2023pilot}, $L_\text{h}+L_\text{c}-1$ symbols at the start of the $N_\text{p}$ length pilot sequence experience interference from the unknown data symbols of the previous frame. To prevent a degradation in the quality of channel estimation, we need to exclude these pilot symbols from any further processing. With that in mind, we focus our processing efforts on the symbols $\tilde{r}_k, k= n_{s_p}^{(i)}+L_\text{h}+L_\text{c}-1,..., {n_{s_p}^{(i)}+N_\text{p}-1}$, which corresponds to the useful segment of the transmitted pilot sequence. The $\textbf{r}_\text{p}= [\tilde{r}_{n_{s_p}^{(i)}+L_\text{h}+L_\text{c}-1}, \tilde{r}_{n_{s_p}^{(i)}+L_\text{h}+L_\text{c}+1}, \ldots, \tilde{r}_{n_{s_p}^{(i)}+N_\text{p}-1}]^\text{T}$ derived from received samples corresponding to the useful part of transmitted pilot sequence in (\ref{equation7}) can be expressed in a vector format as
\begin{IEEEeqnarray}{rcl}\label{equS2Wh}
 \mathbf{r}_\text{p} &{}={}& \mathbf{T} \mathbf{V} \mathbf{c} + \textbf{w}_\text{p},
\end{IEEEeqnarray}
where $\textbf{w}_\text{p}=[w_{n_{s_p}^{(i)}+L_\text{h}+L_\text{c}-1}, w_{n_{s_p}^{(i)}+L_\text{h}+L_\text{c}+1} ..., w_{n_{s_p}^{(i)}+N_\text{p}-1}]^\text{T}$ represents the zero-mean white Gaussian noise with variance $\sigma_n^2$ and $\mathbf{V} \in \mathbb{C}^{(L_\text{h}+L_\text{c}) \times(L_\text{c}+1)}$ is the circulant ISI matrix formed using the elements of vector $\textbf{v}$. The channel coefficients that need to be estimated are denoted by the vector $\textbf{c}$, which is defined as $\textbf{c} = [c_0, c_1, \ldots, c_{L_\text{c}}]^\text{T}$.
The matrix $\mathbf{T} \in \mathbb{C}^{(N_\text{p}-L_\text{h}-L_\text{c}+1)\times (L_\text{h}+L_\text{c})}$ is composed of pilot sequence symbols and can be expressed as introduced in \cite{keykhosravi2023pilot} as follows
\small{
\begin{IEEEeqnarray}{rcl}\label{eqS55noWh}
\mathbf{T} &{}={}& \nonumber\\ 
& & \!\!\! \begin{bmatrix}
s_{n_{s_p}^{(i)}+L_\text{h}+L_\text{c}-1}   & s_{n_{s_p}^{(i)}+L_\text{h}+L_\text{c}-2}     & \dots      &s_{n_{s_p}^{(i)}}\\
s_{n_{s_p}^{(i)}+L_\text{h}+L_\text{c}} & s_{n_{s_p}^{(i)}+L_\text{h}+L_\text{c}-1}        & \dots      &s_{n_{s_p}^{(i)}+1}\\
\vdots   & \vdots   &  \vdots  &\vdots\\
s_{n_{s_p}^{(i)}+N_\text{p}-1} & s_{n_{s_p}^{(i)}+N_\text{p}-2}  & \dots & s_{n_{s_p}^{(i)}+N_\text{p}-L_\text{h}+L_\text{c}}\\
\end{bmatrix}. \nonumber \\
\IEEEeqnarraynumspace
\end{IEEEeqnarray}
} 
\par 
\normalsize
For estimating the unknown channel $\mathbf{c}$ within the useful segment of the pilot sequence (namely, when $k$ lies in the range of $n_{s_p}^{(i)}+L_\text{h}+L_\text{c}$ to $n_{s_p}^{(i)}+N_\text{p}-1$), we employ the LSSE approach, which involves minimizing the sum of squared errors as minimizing the sum of squared errors as
\begin{IEEEeqnarray}{rcl}\label{equS5wh}
\mathbf{\hat{c}} &{}={}& \text{arg} \displaystyle \min_{\mathbf{c}}  (\mathbf{r}_\text{p}-\mathbf{T}   \mathbf{V}   \mathbf{c})^\text{H}(\mathbf{r}_\text{p}-\mathbf{T}   \mathbf{V} \mathbf{c})\nonumber\\
 &{}={}& (\mathbf{TV})^{-1} \mathbf{r}_\text{p}.
\end{IEEEeqnarray}
The matrix $(\mathbf{TV})^{-1}$ can be precomputed offline at the receiver. This is because, the matrix $\mathbf{T}$ is derived from the known pilot sequence, and given the constant taps of the vector $\mathbf{v}$ for an RRC pulse's roll-off factor and the FTN signaling's packing ratio, the matrix $\mathbf{V}$ is also known to the receiver. We choose the pilot sequence in $\mathbf{T}$ that minimizes the MSE for channel estimation in (\ref{equS5wh}) as 
\begin{IEEEeqnarray}{rcl}\label{equS6wh}
\text{MSE}
&{}={}& \text{tr} \left(\sigma_n^2 \mathbf{(TV)}^{-1} \left(\mathbf{(TV)}^{-1}\right)^\text{H}\right).
\end{IEEEeqnarray}
The equation provided above emerges directly from the fact that the estimate $\mathbf{\hat{c}}$ serves as an unbiased estimator of $\mathbf{c}$ in the presence of zero-mean additive Gaussian noise and $\text{E}(\mathbf{\hat{c}}) = \mathbf{c}$. Hence, the MSE of $\mathbf{\hat{c}}$ is equal to its variance. Considering $\text{MSE}$ as a metric assessing the performance of various pilot sequences, an exhaustive search can be conducted to identify the pilot sequence denoted as $\textbf{p} = [s_{n_{s_p}^{(i)}}, s_{n_{s_p}^{(i)}+1}, \ldots, s_{n_{s_p}^{(i)}+N_\text{p}-1}]^\text{T}$ which minimizes the $\text{MSE}$. As represented in \cite{keykhosravi2023pilot}, we have
\begin{IEEEeqnarray}{rcl}\label{eqP2}
\textbf{p} &{}={}& \text{arg} \displaystyle \min_{\textbf{p} \in \mathcal{S}_\text{p}}  \biggl(\text{tr} \left(\sigma_n^2 \mathbf{(TV)}^{-1} \bigl(\mathbf{(TV)}^{-1}\bigr)^\text{H}\right)\biggr).
\end{IEEEeqnarray}
Here, $\mathcal{S}_\text{p}$ represents the constellation set of pilot symbols.

\subsection{Pilot Sequence Location Identification Algorithm}
The proposed pilot sequence location identification algorithm is summarized as:

%%%%%%%%%%%%%%%%%%%%%%%%%%%%%%%%%%
\begin{algorithm}
  \caption{Pilot Sequence Location Identification Algorithm}
  \textbf{Input}:  $\tilde{r}$, $c_1$, and $c_2$\\
  \textbf{Initialize}:  Initialize $\hat{n}_{s_p}$ and $r_{\text{th}_2}$ to zero.\\
  1) Calculate ${|R_{\tilde{p}\tilde{r}} (\delta)|}^2$ according to (\ref{equation34}) for the expected pilot sequence locations $\hat{\mathbb{P}}$.   \\
  2) Sort ${|R_{\tilde{p}\tilde{r}} (\delta)|}^2$ for the expected pilot sequence locations, $\hat{\mathbb{P}}$, in descending order and find the maximum value $m$. \\
  3) Select the candidate locations from the set $\hat{\mathbb{P}}$, for which ${|R_{\tilde{p}\tilde{r}} (\delta)|}^2$ exceeds $r_{\text{th}_1}=c_1 \times m$. \\
  4) \textbf{For} each candidate \textbf{do} \\
       \hspace*{2.5em} Find the corresponding $\delta_{c_0}$ and $\delta_{c_{L_c}}$. \\
        \hspace*{2.5em} Calculate the measure, $\mu(\delta_{c_0},\delta_{c_{L_c}})$ according to (\ref{equation36}).\\
        \hspace*{2.5em} \textbf{If} the calculated $\mu(\delta_{c_0},\delta_{c_{L_c}})>r_{\text{th}_2}$, \textbf{then}, \\
        \hspace*{5em} update the $r_{\text{th}_2}=c_2 \times \mu(\delta_{c_0},\delta_{c_{L_c}})$ and $\hat{n}_{s_p}=\delta_{c_0}$\\ 
        \hspace*{2.5em} \textbf{EndIf}\\
    \hspace*{1em} \textbf{EndFor}
  
  \textbf{Output}:  $\hat{n}_{s_p}$
\end{algorithm}

\begin{table*}[t]
  \centering
  \caption{{Complexity comparison of the PSLI algorithm and channel estimation in this paper with the works in \cite{ishihara2017iterative} and \cite{keykhosravi2023pilot}.} }
  \begin{tabular}{|c|c|}
    \hline
    Algorithm & Computational complexity \\
    \hline
    MMSE-based channel estimation for FTN signaling system in \cite{ishihara2017iterative}&  $\mathcal{O}\Big({N_\text{p} \log}(N_\text{p}) + N_\text{p}\Big)$ \\
    \hline
    LSSE-based  channel estimation for FTN signaling system in \cite{keykhosravi2023pilot}&
    $\mathcal{O}\Big( (N_\text{p}-L_\text{h}-L_\text{c}+1) \cdot (L_\text{c}+1) \Big)$\\
    \hline
    & 
    $\mathcal{O}\Big( (N_\text{p}-L_\text{h}-L_\text{c}+1) \cdot (L_\text{c}+1) + $ \\
    PSLI algorithm and LSSE-based IM-based channel estimation for FTN signaling system 
    &$ N_\text{p} \times \text{Card}\{\hat{\mathbb{P}}\} + \text{Card}\{\hat{\mathbb{P}}\} \times {\log}{\big(\text{Card}\{\hat{\mathbb{P}}\}\big)} $\\
    & $ +  {\log}{\big(\text{Card}\{\hat{\mathbb{P}}\} \big)} \small{\times 4\times \text{Card}\{\hat{\mathbb{P}}\}} \normalsize{} \times  L_\text{c}\Big)$
  \\
    \hline
    \end{tabular}
  \label{tab:table_2}
\end{table*}

\subsection{Complexity Analysis}

%%%%%%%%%%%%%%%%%%%%%%%%%%%%%%%%
This subsection presents a complexity analysis of the proposed PSLI algorithm. In the first step, we calculate ${|R_{\tilde{p}\tilde{r}} (\delta)|}^2$ using (\ref{equation34}) for the $\hat{\mathbb{P}}$ locations. The computational complexity of this step is $\mathcal{O}\big(N_\text{p} \cdot \text{Card}\{\hat{\mathbb{P}}\}\big)$. 
To sort the ${|R_{\tilde{p}\tilde{r}} (\delta)|}^2$ for the expected pilot sequence locations, $\hat{\mathbb{P}}$, in the second step, one requires a complexity of $\mathcal{O}\big(\text{Card}\{\hat{\mathbb{P}}\}\cdot {\log}{(\text{Card}\{\hat{\mathbb{P}}\}}\big)$. As in the third step the ${|R_{\tilde{p}\tilde{r}} (\delta)|}^2$ values for the  $\hat{\mathbb{P}}$ locations are sorted, finding the location for which ${|R_{\tilde{p}\tilde{r}} (\delta)|}^2 > r_{\text{th}_1}$ requires $\mathcal{O}\big({\log}{(\text{Card}\{\hat{\mathbb{P}}\})}\big)$ computational complexity. In the fourth step, first, we need to calculate the congruence modulo to find $\delta_{c_0}$ and $\delta_{c_{L_c}}$. This step has a constant computational complexity. Secondly, calculating the $\mu(\delta_{c_0},\delta_{c_{L_c}})$ according to (\ref{equation36}) involves calculation of $\lambda (\delta_{c_0})$ and $\lambda (\delta_{c_{L_c}})$ according to (\ref{equation32_1}) which requires summation over $2r_0$ elements where $r_0< L_\text{c}$. Thus, one requires $\mathcal{O}( 4\cdot L_\text{c})$. The third part in the fourth step has a constant computational complexity. Thus, the fourth step requires the complexity of $\mathcal{O}\big( 4\cdot \text{Card}\{\hat{\mathbb{P}}\} \cdot L_\text{c}\big)$. Having said that, the total computational complexity of the proposed PSLI algorithm can be written as $\begin{aligned}
&\left. \! \small{\mathcal{O}\Big(N_\text{p} \times \text{Card}\{\hat{\mathbb{P}}\} + \text{Card}\{\hat{\mathbb{P}}\} \times {\log}{\big(\text{Card}\{\hat{\mathbb{P}}\}\big)} +  {\log}{\big(\text{Card}\{\hat{\mathbb{P}}\}\big)}}\right. \\
&\left.
\small{\times 4\times \text{Card}\{\hat{\mathbb{P}}\}} \normalsize{} \times  L_\text{c}\Big) \text{, where we have $\text{Card}\{\hat{\mathbb{P}}\}$ is equal to } \right. \\
&\left. 2^{{\lfloor {\log_2}{(N_d+1)} \rfloor}-\lceil{{\log_2}{(L_\text{c}+1)}}\rceil+1}\text{.}\right.
\end{aligned}$

{The complexity of channel estimation can be calculated from (\ref{equS5wh}).
As mentioned before, the matrix $(\mathbf{TV})^{-1}$ can be precomputed offline at the receiver. 
Having said that, the computational complexity of (\ref{equS5wh}) is $\mathcal{O}\big( (N_\text{p}-L_\text{h}-L_\text{c}+1) \cdot (L_\text{c}+1) \}\big)$. This is the same as the complexity of channel estimation in \cite{keykhosravi2023pilot}.}

{The computational complexity of the proposed channel estimation in \cite{ishihara2017iterative}
can be computed based on (32) and (35) in their work. The channel estimation in (32) of \cite{ishihara2017iterative} requires, FFT and inverse FFT operation and diagonal matrix multiplication as well as calculation of minimum mean squared error (MMSE) weights matrix according to (35) in \cite{ishihara2017iterative}. Thus the computational complexity of the channel estimation in \cite{ishihara2017iterative} is $\mathcal{O}\big({N_\text{p} \log}(N_\text{p}) \big)+\mathcal{O}\big(N_\text{p}\big)$.}

{Table \ref{tab:table_2} summarizes the computational complexity of the PSLI algorithm and channel estimation in this paper compared to the channel estimation methods in  \cite{keykhosravi2023pilot} and \cite{ishihara2017iterative}. It should be noted as the pilot location is fixed in \cite{keykhosravi2023pilot} and \cite{ishihara2017iterative}, hence there is no need to employ an algorithm to identify the location of the pilot in these papers.
}

\subsection{Pilot Sequence Location Identification for Doubly-selective HF Channel}

In the context of time-invariant channels, a single estimation of the channel suffices for the entire frame. More specifically, once the pilot sequence is employed to estimate the channel within a given frame, that estimation remains consistent throughout the entire data block of that frame. Conversely, when dealing with doubly-selective channels, the initially estimated channel at the pilot sequence location in the frame's beginning may become outdated as the frame progresses. Consequently, this can lead to an accumulation of mean squared error over the duration of the frame. To mitigate the inaccuracies arising from assuming identical channel coefficients across the entire frame, an interpolation technique can be employed. 
By utilizing these techniques, the coefficients can be estimated by interpolating between the pilot sequence of the current frame and the pilot sequence of the next frame. Performing interpolation alleviates the impact of outdated channel estimates, thereby enhancing the accuracy of the estimation process.

The investigation conducted in \cite{keykhosravi2023pilot} explores the impact of three interpolation methods on the MSE associated with estimating the doubly-selective channel. The study reveals that the MSE remains relatively consistent across the various interpolation techniques employed, namely linear, cubic, and spline interpolation. That being stated, we opt for the utilization of linear interpolation due to its inherent advantage of low computational complexity.

Let the channel estimation at the pilot sequence of the current frame, i.e., $i$th frame, and the pilot sequence of the next frame, i.e., $(i+1)$th frame be $\mathbf{\hat{c}_i}$ and $\mathbf{\hat{c}_{i+1}}$, respectively. To update the channel estimation at each data symbol $k$ within the $i$th frame, where $k = 0,...,{{N_\text{s}}}-1$, we employ a linear interpolation technique that leverages the values  {$\mathbf{\hat{c}_i}$} and {$\mathbf{\hat{c}_{i+1}}$}.

\begin{IEEEeqnarray}{rcl}\label{equInt}
\mathbf{\hat{c}}_{i,k+1} &{}={}& \mathbf{\hat{c}}_{i,k} + 
\frac{(\mathbf{\hat{c}_{i+1}}-\mathbf{\hat{c}_{i}})}{{{N_\text{s}}}}, \quad k=0,...,{{N_\text{s}}}-1, \IEEEeqnarraynumspace
\end{IEEEeqnarray}
where {$\mathbf{\hat{c}}_{i,k}$} denote the channel estimation for the $k$th data symbol within the $i$th frame. Additionally, {$\mathbf{\hat{c}}_{i,0}$} is initialized as $\mathbf{\hat{c}_i}$, representing the channel estimation at the pilot sequence of the current frame.

\section{Simulation Results}

In this section, numerical results of the performance evaluation of the proposed IM-based channel estimation algorithm in terms of pilot sequence location identification error (PSLIE), MSE and {BER} are presented.
\subsection{{Simulation Setup}}
We select a typical symbol rate of 2400 symbols/sec on the transmitter side. {We employ a rate $1/2$ convolutional code characterized by a constraint length of $7$ and utilizing the two generator polynomials 0x5b and 0x79.
By puncturing the bits of a rate 1/2 code with a puncturing mask of $[1, 1, 1, 0, 0, 1]$, a code rate of $R_\text{c} = 3/4$ can be attained. In this context, a “1” signifies that the corresponding bit is transmitted, while a “0” implies that the bit is discarded. 
At the receiver, a decoding algorithm based on a posteriori probability (APP) is utilized to decode the convolutional code. Binary phase shift keying (BPSK) and quadrature phase shift keying (QPSK) modulations are used for pilot and data symbols, respectively unless explicitly specified otherwise. 
We conducted an exhaustive search to identify the pilot sequence, $\textbf{p}$,
which minimizes the MSE as outlined in (\ref{equS6wh}). It's worth noting that, the optimal pilot sequence is determined by implementing (\ref{eqP2}).}

In our MSE performance simulations described in this paper, we employed an exhaustive search to {find} the optimal pilot sequence for minimizing the MSE as outlined in (\ref{eqP2}). This choice was made primarily because pilot sequence design is typically a one-time, offline task for a given channel. However, for pilot sequences of higher modulation order, the exhaustive search process becomes impractical. % due to its time-consuming nature. 
Therefore, we opted for a sub-optimal approach to tackle this issue. This sub-optimal solution involves relaxing the integer constraint of (\ref{eqP2}) and solving the relaxed optimization problem corresponding to (\ref{eqP2}) numerically using the interior-point algorithm \cite{wright1997primal}. The ultimate pilot sequence is then determined by rounding this sub-optimal solution. It's important to note that we can no longer assure optimality as a result of this rounding procedure over the continuous suboptimal. It's important to note that since the optimization problem is non-convex, the interior-point algorithm may converge to a local optimal solution, which may not necessarily be the global optimal solution. To enhance the likelihood of obtaining a high-quality local sub-optimal solution, we initialized the interior-point algorithm with uniformly distributed random pilot sequences and repeated the simulation 100 times. We then computed the MSE for each of these sequences and selected the sub-optimal pilot sequence with the lowest MSE value.
Having said that we demonstrate the effectiveness of our algorithm with higher-order modulations by employing 8-PSK modulation, which is utilized in HF transmissions, for both pilot and data symbols.
An RRC pulse shaping filter having a roll-off factor of $\beta = 0.35$ is employed unless stated otherwise.

%%%%%%%%%%%%%%%%%%%%%%%
\begin{figure}[t]
\centering
\includegraphics[width=8.5cm,height=6cm]{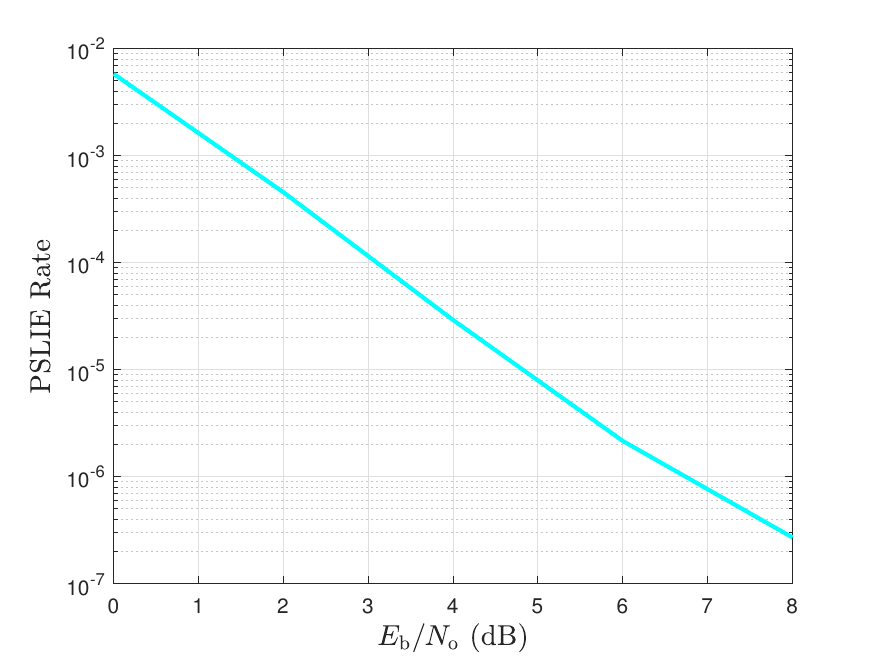}
\caption{The PSLIE rate of the proposed IM-based channel estimation algorithm for FTN signaling with $\tau = 0.8$ over \textit{channel model 3}.}
\label{fig:7}
\end{figure}

\begin{figure}[t]
\centering
\includegraphics[width=8.5cm,height=6cm]{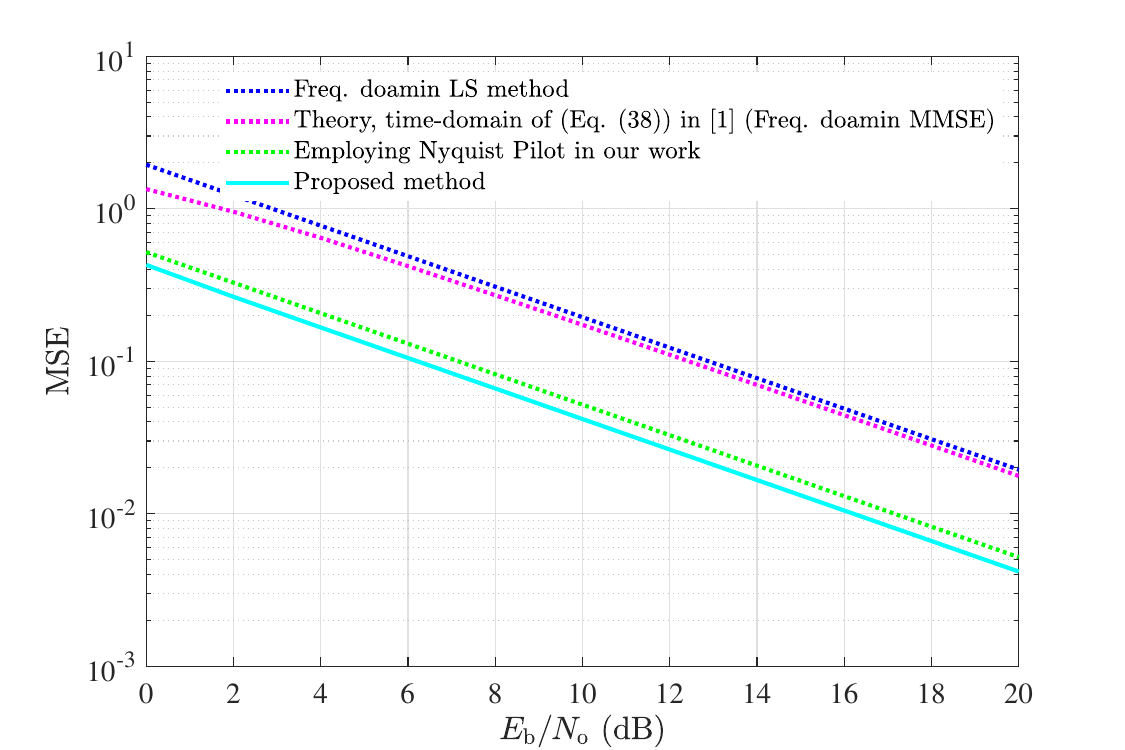} %MSEtau8
\caption{{The MSE of the proposed FTN the proposed IM-based channel estimation
algorithm for FTN signaling employing the designed optimal pilot sequence versus the methods %in \cite{ishihara2017iterative} 
in the literature over \textit{channel model 3} for $\tau = 0.8$.}}
\label{fig:8}
\end{figure}
%%%%%%%%%%%%%%%%%%%
\begin{figure}[t]
\centering
\includegraphics[width=8.5cm,height=6cm]{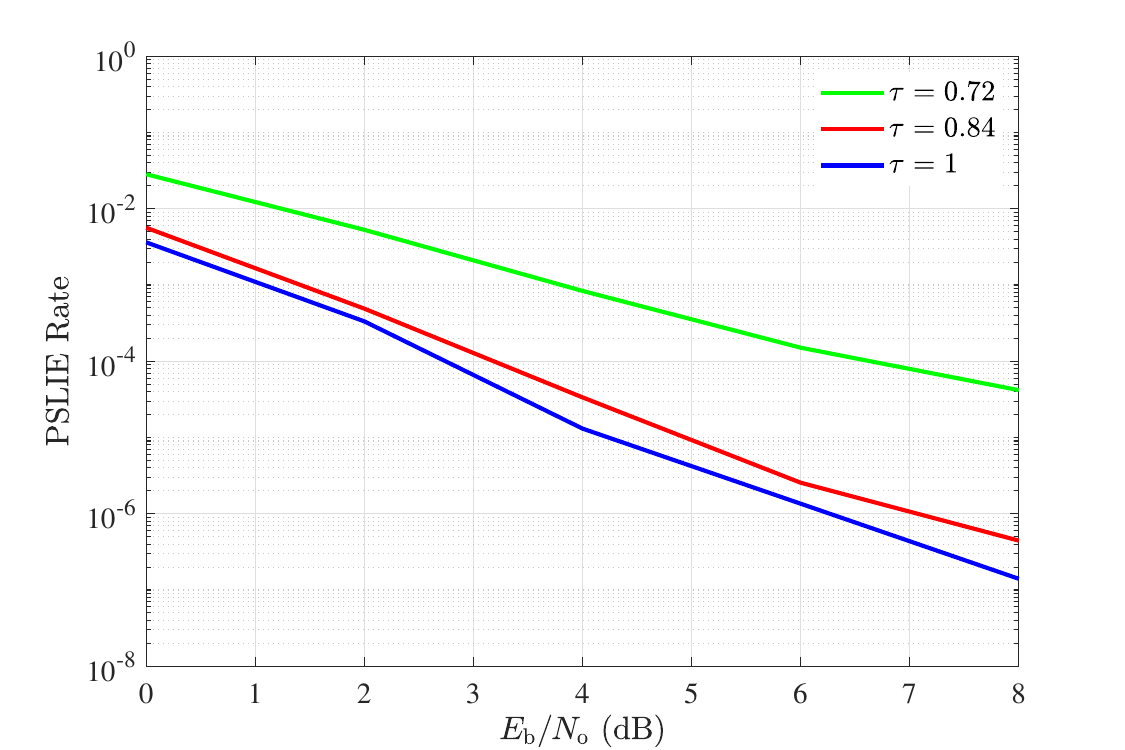}
\caption{The PSLIE rate of the proposed IM-based channel estimation algorithm for FTN signaling with $\tau = 0.72$, $\tau = 0.84$ {, and $\tau = 1$} over \textit{channel model 2}.}
\label{fig:9}
\end{figure}

\begin{figure}[t]
\centering
\includegraphics[width=8.5cm,height=6cm]{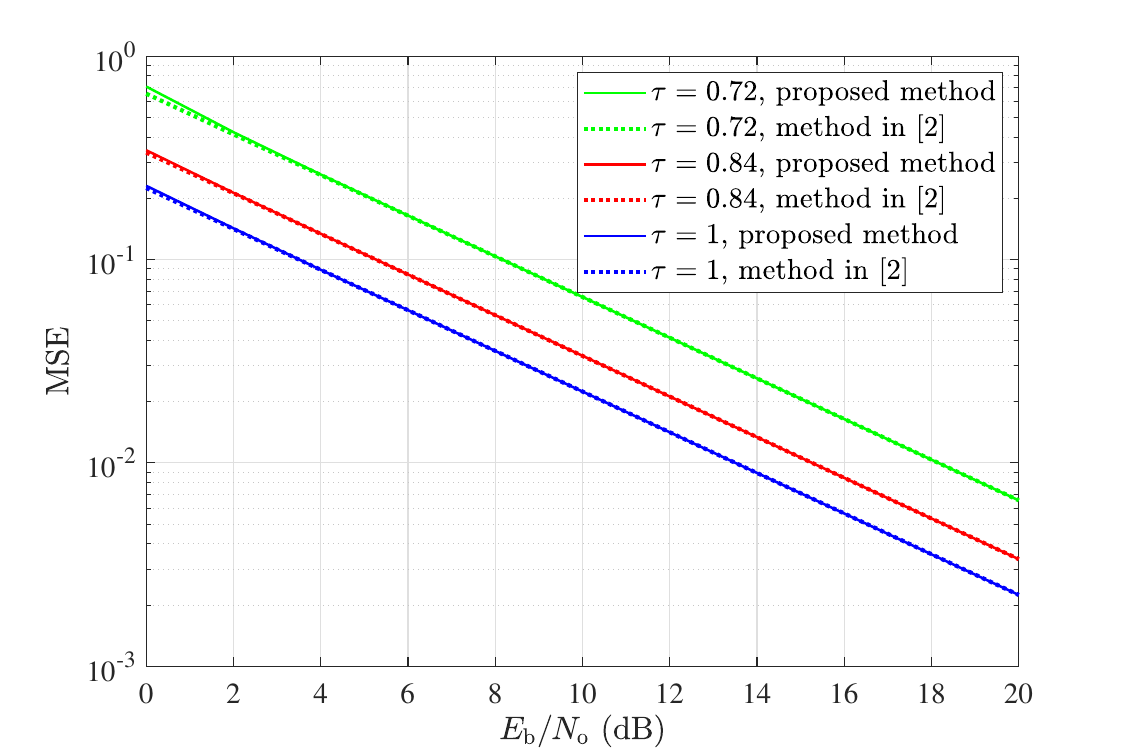}
\caption{The MSE of the proposed IM-based channel estimation algorithm for FTN signaling with $\tau = 0.72$, $\tau = 0.84$, {and $\tau = 1$} over \textit{channel model 2}.}
\label{fig:10}
\end{figure}
%%%%%%%%%%%%%%%%%%%%
\begin{figure}[t]
\centering
\includegraphics[width=8.5cm,height=6cm]{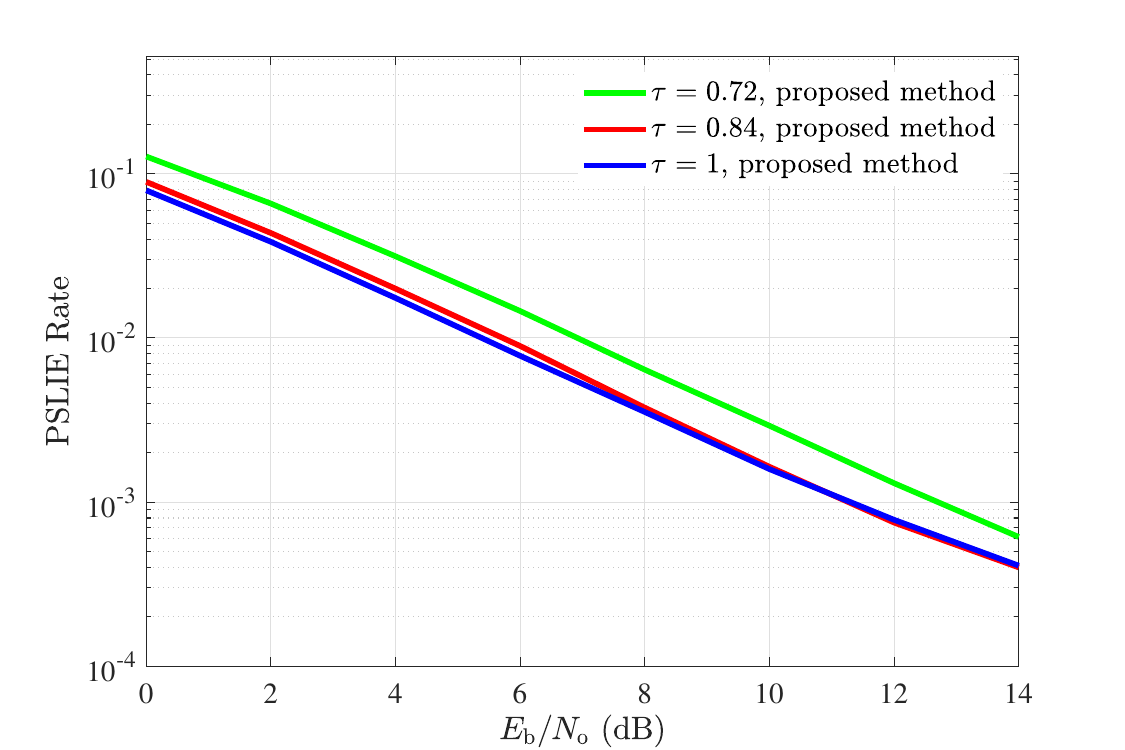}
\caption{The PSLIE rate of the proposed IM-based channel estimation algorithm for FTN signaling with {$\tau = 0.72$, $\tau = 0.84$, and $\tau = 1$} over \textit{channel model 1}.}
\label{fig:PSLI_HF}
\end{figure}

\begin{figure}[t]
\centering
\includegraphics[width=8.5cm,height=6cm]{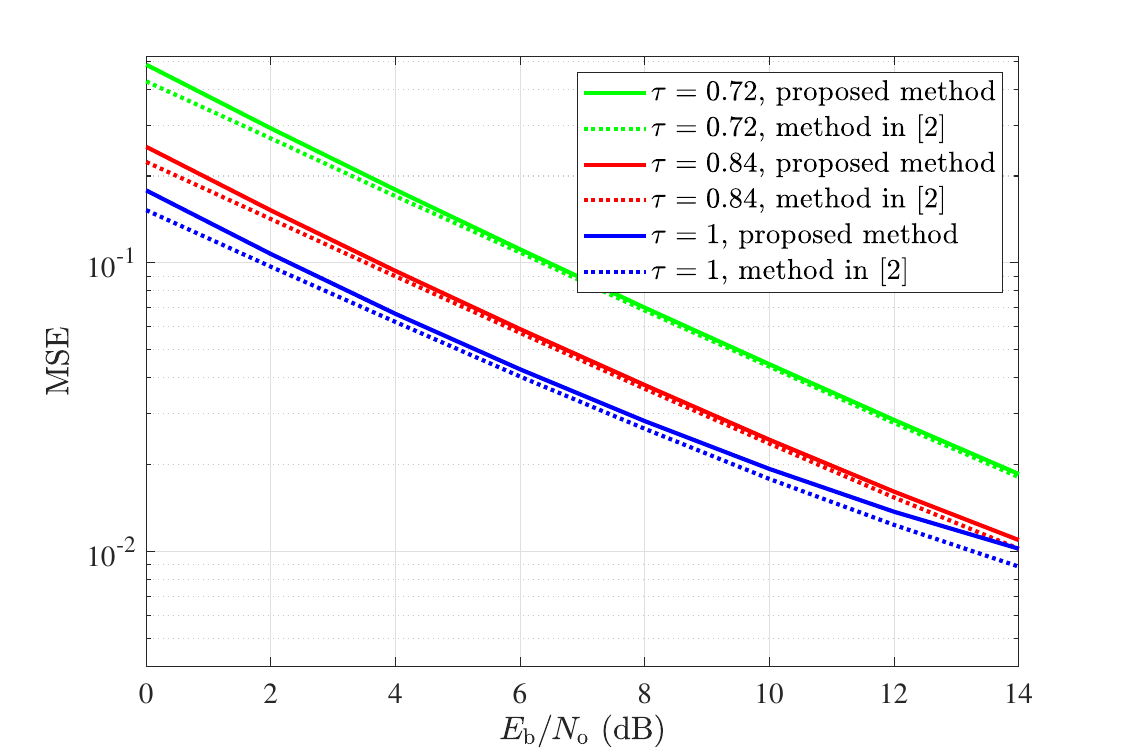}
\caption{The MSE of the proposed IM-based channel estimation algorithm for FTN signaling with {$\tau = 0.72$, $\tau = 0.84$, and $\tau = 1$} over \textit{channel model 1}.}
\label{fig:12}
\end{figure}
%%%%%%%%%%%%%%%%%%%%%%%%%%%%%%%%%
\begin{figure}[t]
\centering
\includegraphics[width=8.5cm,height=6cm]{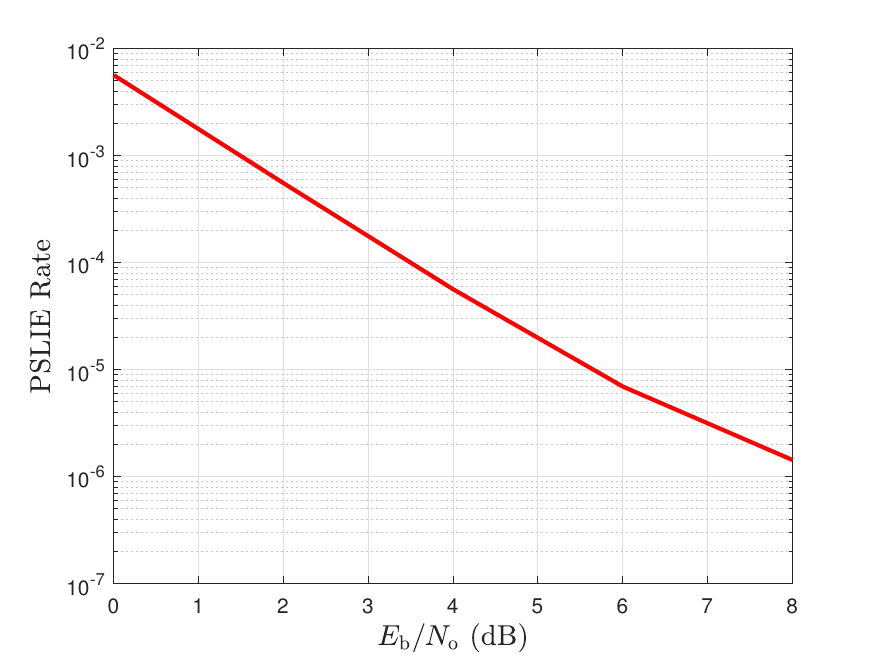}
\caption{The PSLIE rate of the proposed IM-based channel estimation algorithm for FTN signaling with $\tau = 0.84$ over \textit{channel model 2} where pilot and data symbols are modulated with 8-PSK.}
\label{fig:21}
\end{figure}

\begin{figure}[t]
\centering
\includegraphics[width=8.5cm,height=6cm]{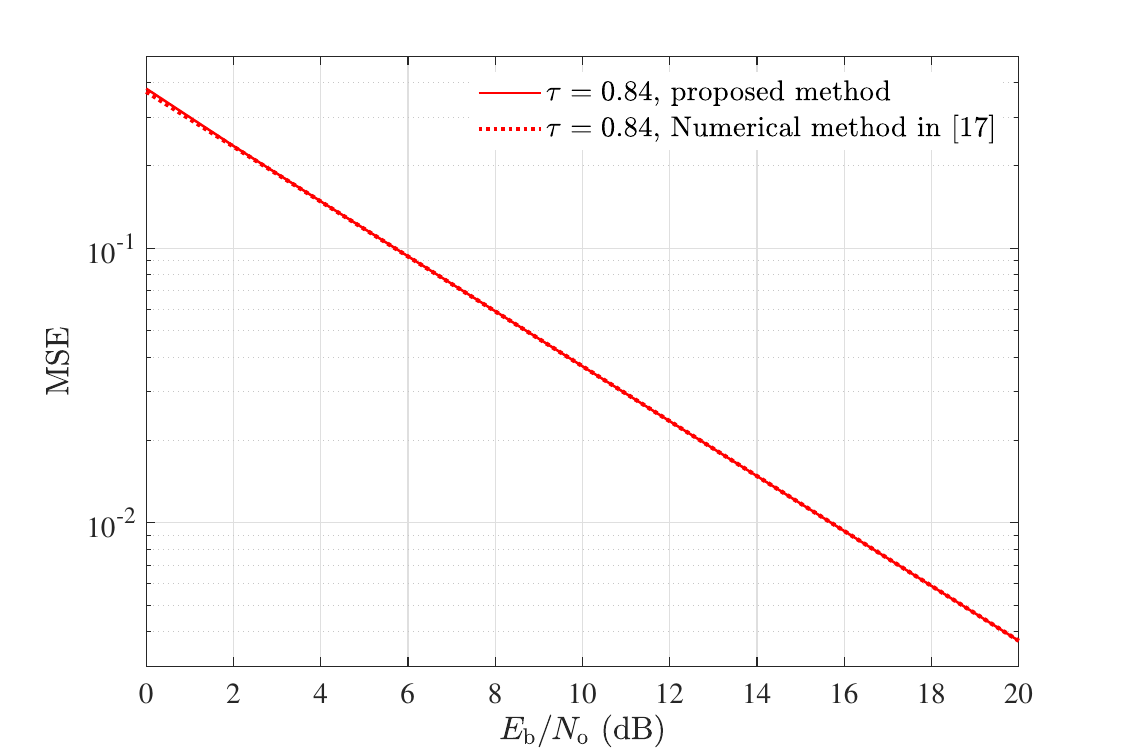}
\caption{The MSE of the proposed IM-based channel estimation algorithm for FTN signaling with $\tau = 0.84$ over \textit{channel model 2} where pilot and data symbols are modulated with 8-PSK. The pilot sequence was found using numerical optimization.}
\label{fig:20}
\end{figure}
%%%%%%%%%%%%%%%%%%%%%%%%%%%%%%%%%%%%%%%
We investigate three distinct channel models in our study. Firstly, we examine the ITU-R Poor channel, which is characterized as a doubly-selective Rayleigh fading channel. This channel exhibits a fading rate of $0.0004$ (Doppler frequency of 1 Hz) and a delay spread of 2.1 ms \cite{ITU}. We refer to it as \textit{channel model 1}. We set the values of $N_\text{p}$ and ${{N_\text{s}}}$ to 32 and 256 symbols, respectively.
Secondly, we consider a frequency-selective Rayleigh fading channel with a delay spread similar to that of \textit{channel model 1}. This particular channel, referred to as \textit{channel model 2} which can be regarded as the time-invariant version of channel model 1. Lastly, we examine a frequency-selective Rayleigh fading channel denoted as \textit{channel model 3}. It possesses a channel of $L_\text{c} = 7$ and maintains a constant impulse response over two successive transmission frames. It is important to note that the inclusion of \textit{channel model 3} primarily serves the purpose of facilitating a fair comparison with the work presented in \cite{ishihara2017iterative}.
%%%%%%%%%%%%%%%%%%%%%

For each run of the simulation, a superframe comprising 144 frames is transmitted through the channel. The PSLIE rate represents the percentage of packets in which the receiver fails to correctly identify the pilot sequence location. The MSE is computed for each individual simulation run as $\sum\nolimits_{l=0}^{L_{c}} ({\hat{\textbf{c}}_l}-\textbf{c}_l)^\text{H}({\hat{\textbf{c}}_l}-\textbf{c}_l)$, and the final simulated MSE is obtained by taking the average of the MSE values calculated across all simulation runs.

\subsection{{PSLIE and MSE results}}
%%%%%%%%%%%%%%%%%%%%%
Fig. \ref{fig:7} depicts the PSLIE of the proposed IM-based channel estimation algorithm for FTN signaling with $\tau = 0.8$ over \textit{channel model 3}. As shown in Fig. \ref{fig:7}, increasing the $E_\text{b}/N_0$ leads to a decrease in PSLIE. The PSLIE for $E_\text{b}/N_0$ as low as 3 dB is shown to be about $10^{-4}$. As $E_\text{b}/N_0$ increases to 8 dB, the PSLIE becomes even lower, reaching approximately $2\times 10^{-7}$. This striking result clearly demonstrates that the proposed IM-based channel estimation algorithm enables precise identification of the pilot sequence location.

Fig. \ref{fig:8} presents a comparison of the MSE of the proposed IM-based channel estimation algorithm for FTN signaling with $\tau = 0.8$ over \textit{channel model 3} with the MSE of the work in \cite{ishihara2017iterative} having $\beta = 0.5$. {The MSE for the case of using the Nyquist pilot sequence instead of the optimal pilot sequence designed for channel estimation over FTN signaling in our work and the case of using the LS method in the frame structure model of \cite{ishihara2017iterative} is illustrated in Fig. \ref{fig:8}.
It should be noted that in all these cases except our proposed work illustrated in Fig. \ref{fig:8}, the pilot sequence location is fixed in the frame, and thus no additional information bits are conveyed through the location of the pilot.}
In order to maintain equal overhead between our work and the work in \cite{ishihara2017iterative}, the pilot sequence length in Ishihara et al.'s work \cite{ishihara2017iterative} is set to 8. This choice is made because, in their approach, a two-pilot sequence length is added at the beginning and end of each data block to form a frame. By setting the pilot length to 8 in \cite{ishihara2017iterative}, they achieve a total of 32 known symbols (overhead of 32) in each frame, which is equivalent to the overhead in a given frame in our paper. This consistency of overhead ensures a fair and consistent comparison between the two methods in terms of their pilot usage and overhead. The MSE in Ishihara et al.'s work \cite{ishihara2017iterative} is formulated in the frequency domain, as indicated in Eq. (38) of their paper. To ensure a meaningful and equitable comparison of MSE with our own work, we perform a conversion of the MSE expression from the frequency domain to the time domain. This conversion allows for a consistent assessment of MSE between the two approaches, despite their different domain representations, and facilitates a comprehensive evaluation of the performance of both methods in a unified framework. {The work in \cite{ishihara2017iterative} is based on the MMSE method to estimate the channel. We used the LS method in the frame structure introduced in \cite{ishihara2017iterative} to find the optimal pilot and estimate the channel for FTN signaling in the frequency domain. We also compared our work with the case when the Nyquist pilot sequence (32 which is of the same length as our work) is employed instead of the optimal pilot sequence designed for channel estimation over FTN signaling.} 
A noteworthy point is that the time-domain MSE can be equivalent to calculating $\sum\nolimits_{l=0}^{L_{c}} ({\hat{\textbf{c}}_l}-\textbf{c}_l)^\text{H}({\hat{\textbf{c}}_l}-\textbf{c}_l)$. As depicted in Fig. \ref{fig:8}, the proposed IM-based channel estimation algorithm in this paper results in a significantly better (about 6 dB) MSE when compared to the MSE resulting from the channel estimation method in \cite{ishihara2017iterative}. The SE in our work can be written as $\text{SE} = {\frac{{{N_\text{s}}}\cdot{{R_\text{c}}}\cdot{{\log_2}M}+{N_\text{b}}}{({N_\text{s}}+N_\text{p})} \cdot\frac{1}{\tau (1+\beta)} = \frac{256{\cdot \frac{3}{4}}\cdot {\log_2}4+5}{(256+32)} \cdot \frac{1}{0.8(1+0.5)}   = 1.1256}$ bits/s/Hz for $\tau = 0.8$. For the same value of $\tau = 0.8$, the SE in \cite{ishihara2017iterative} can be written as $\text{SE} = \frac{{{N_\text{s}}}}{({{N_\text{s}}}+N_\text{p})} \cdot\frac{{\log_2}M}{\tau (1+\beta)} {{R_\text{c}}} = \frac{256}{(256+32)} \cdot \frac{{\log_2}4}{0.8(1+0.5)} {\cdot {\frac{3}{4}}}= {{1.1111}}$ bits/s/Hz. That is a {1.3\%} improvement in SE when utilizing the proposed IM-based channel estimation as well as improvement {(about 6 dB)} in the MSE.
{As simulation results in Fig. \ref{fig:8} show, the proposed IM-based channel estimation exhibits more than a 6 dB improvement in MSE compared to the LS channel estimation in the frequency domain while gaining SE improvement. Additionally, there is approximately 0.5 dB improvement when using the optimal pilot sequence for the proposed IM-based channel estimation of FTN signaling compared to the case of using the Nyquist pilot while achieving SE improvement.}
%%%%%%%%%%%%%%%%%%%%%%%%%%%%

{Fig. \ref{fig:9} portrays the PSLIE of the proposed IM-based channel estimation algorithm for FTN signaling with $\tau = 0.72$, $\tau = 0.84$, and $\tau = 1$ over \textit{channel model 2}. As expected, the PSLIE for $\tau = 0.84$ is lower compared to $\tau = 0.72$. %The PSLIE for $\tau = 0.84$ and $\tau = 1$ remain comparable while PSLIE for $\tau = 1$ is slightly lower than $\tau = 0.84$. 
As evidenced by Fig. \ref{fig:9}, an increase in the $E_\text{b}/N_0$ results in a corresponding reduction in the PSLIE. 
For $E_\text{b}/N_0$ as low as 4 dB, considering $\tau = 0.72$, $\tau = 0.84$, and $\tau = 1$, the PSLIE stands approximately at $10^{-3}$, $3 \times 10^{-5}$ and $1.3 \times 10^{-5}$, respectively. The PSLIE continues to decrease as the $E_\text{b}/N_0$ is increased. Notably, for $E_\text{b}/N_0$ of 8 dB, the PSLIE reaches approximately $4\times 10^{-5}$, $4.5\times 10^{-7}$ and $1.4 \times 10^{-7}$ for $\tau = 0.72$, $\tau = 0.84$, and $\tau =1$, respectively. This remarkable outcome shows the efficiency of our proposed algorithm, enabling precise and accurate identification of pilot sequence locations within the transmitted frames.}

Fig. \ref{fig:10} offers comparative results of the MSE for our proposed IM-based channel estimation algorithm for FTN signaling with $\tau = 0.72$, $\tau = 0.84${, and $\tau = 1$} over \textit{channel model 2} with \cite{keykhosravi2023pilot}. As clearly illustrated in Fig. \ref{fig:10}, the MSE maintains consistent with the MSE results presented in \cite{keykhosravi2023pilot}, even for $E_\text{b}/N_0$ as low as 2 dB.
{As shown in Table \ref{tab:table_1},} the SE employed in our proposed IM-based channel estimation for the FTN signaling having the code rate of $3/4$ equals {$1.3896$} bits/s/Hz for $\tau = 0.72$.
The SE in \cite{keykhosravi2023pilot} for the same value of $\tau$ amounts to {1.3717} bits/sec/Hz, indicating about {1.3\%} improvement in SE when utilizing the proposed IM-based channel estimation in the HF channel while maintaining the same MSE as \cite{keykhosravi2023pilot}.
{The SE for the traditional Nyquist case is equal to $0.9877$. This indicates a {$40.70\%$} improvement in SE when utilizing the proposed IM-based channel estimation for the FTN signaling compared to the traditional Nyquist signaling.}

%%%%%%%%%%%%%%%%%%%%%%%%%%%%%%%%%%%%%%%

{In Fig. \ref{fig:PSLI_HF}, the PSLIE of the proposed IM-based channel estimation algorithm of doubly-selective HF channel (\textit{channel model 1}) for FTN signaling with $\tau = 0.72$, $\tau = 0.84$, and $\tau = 1$ is presented. 
For $E_\text{b}/N_0$ of 10 dB, the PSLIE is observed to be less than $3\times 10^{-3}$ for $\tau = 0.72$ and less than $2\times 10^{-3}$ for $\tau = 0.84$, and $\tau = 1$ cases. For the aforementioned values of $\tau$, as $E_\text{b}/N_0$ increased to 12 dB, the PSLIE experiences a further reduction, approaching or falling below a value of $10^{-3}$. This outcome underscores the capability of the proposed algorithm to achieve precise identification of the pilot sequence location within the transmitted frames.}

{Fig. \ref{fig:12} provides a comparative examination of the MSE for our proposed IM-based channel estimation algorithm for FTN signaling with $\tau = 0.72$, $\tau = 0.84$, and $\tau = 1$ over \textit{channel model 1} with \cite{keykhosravi2023pilot}. For the three values of $\tau$, Fig. \ref{fig:12} reveals a degradation less than 0.5 dB for the $E_\text{b}/N_0$ of 14 dB for the proposed IM-based channel estimation when compared to the MSE in \cite{keykhosravi2023pilot}. Thus, the increase in SE in this scenario comes at the price of degradation in the MSE.}

The PSLIE for the proposed IM-based channel estimation algorithm of FTN signaling with $\tau = 0.84$ over \textit{channel model 2} is illustrated in Fig. \ref{fig:21}. In this scenario, 8-PSK modulation is employed for both the pilot sequence and data symbols. As depicted in Fig. \ref{fig:21}, an increase in $E_\text{b}/N_0$ corresponds to a reduction in PSLIE. Notably, even at a relatively low $E_\text{b}/N_0$ value of 4 dB, the PSLIE is as low as $5 \times 10^{-5}$. With the $E_\text{b}/N_0$ increasing to 8 dB, the PSLIE exhibits an even more significant reduction, reaching approximately $10^{-6}$. This impressive result unequivocally attests to the precision enabled by the proposed IM-based channel estimation algorithm in identifying the pilot sequence location for higher-order modulations for both data and pilot symbols within the frames.

In Fig. \ref{fig:20}, a comparison is shown between the MSE of the proposed IM-based channel estimation for FTN signaling with $\tau = 0.84$ over \textit{channel model 2} with its counterpart in \cite{keykhosravi2023pilot}. To illustrate the efficiency of when dealing with higher-order modulations, we have adopted 8-PSK modulation, which is commonly employed in HF transmissions, for both pilot and data symbols.
Fig. \ref{fig:20} demonstrates that, even with a low $E_\text{b}/N_0$ value of 2 dB, the MSE consistently matches the MSE reported in \cite{keykhosravi2023pilot}. For the case of our proposed IM-based channel estimation for the FTN signaling, the SE as defined in (\ref{equationSE}) equals to ${\gamma_{\text{IM-FTN}} = 1.7790}$ bits/s/Hz when $\tau = 0.84$. 
In \cite{keykhosravi2023pilot}, the SE for the same value of $\tau$ is reported as {1.7637} bits/sec/Hz. This demonstrates a {0.87} percent enhancement in SE when employing the proposed IM-based channel estimation in the HF channel, a spectrum-scarce environment. This improvement in SE is achieved while keeping the MSE at the same level as reported in \cite{keykhosravi2023pilot}.

\begin{figure}[t]
\centering
\includegraphics[width=8.5cm,height=6cm]{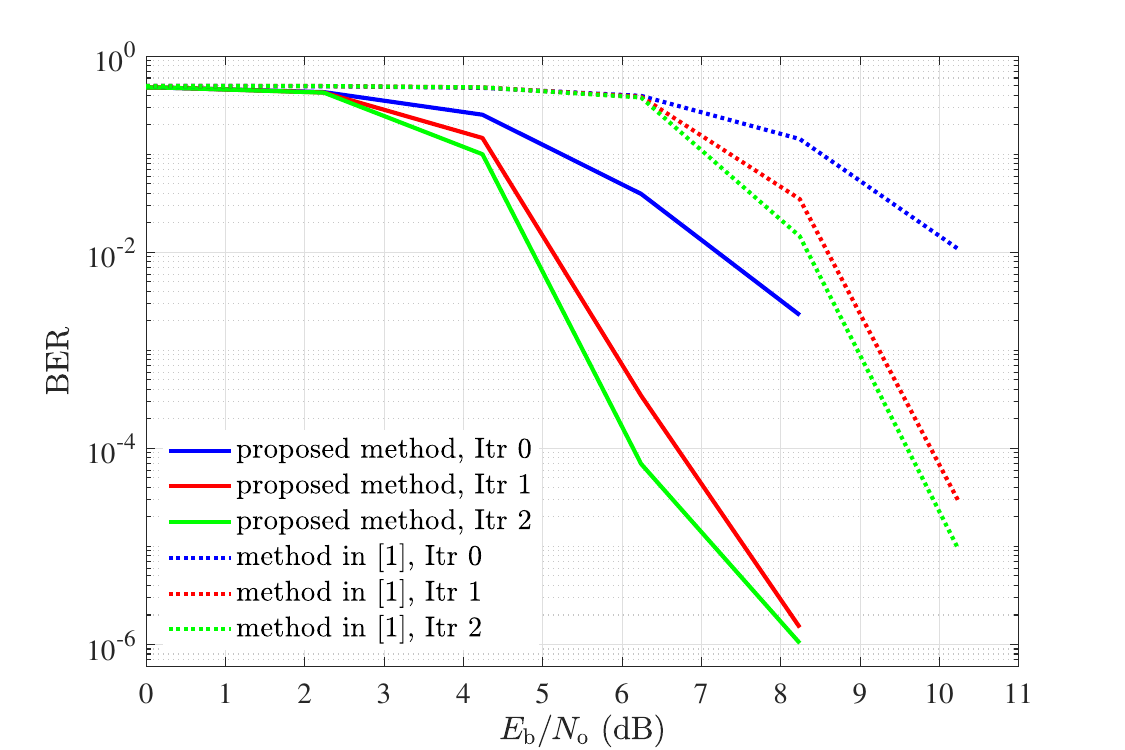}
\caption{{The BER performance of the proposed IM-based channel estimation
algorithm for FTN signaling employing the designed optimal pilot sequence versus the method in \cite{ishihara2017iterative} over \textit{channel model 3} for $\tau = 0.8$.}}
\label{fig:BER_2017}
\end{figure}

\begin{figure}[t]
\centering
\includegraphics[width=8.5cm,height=6cm]{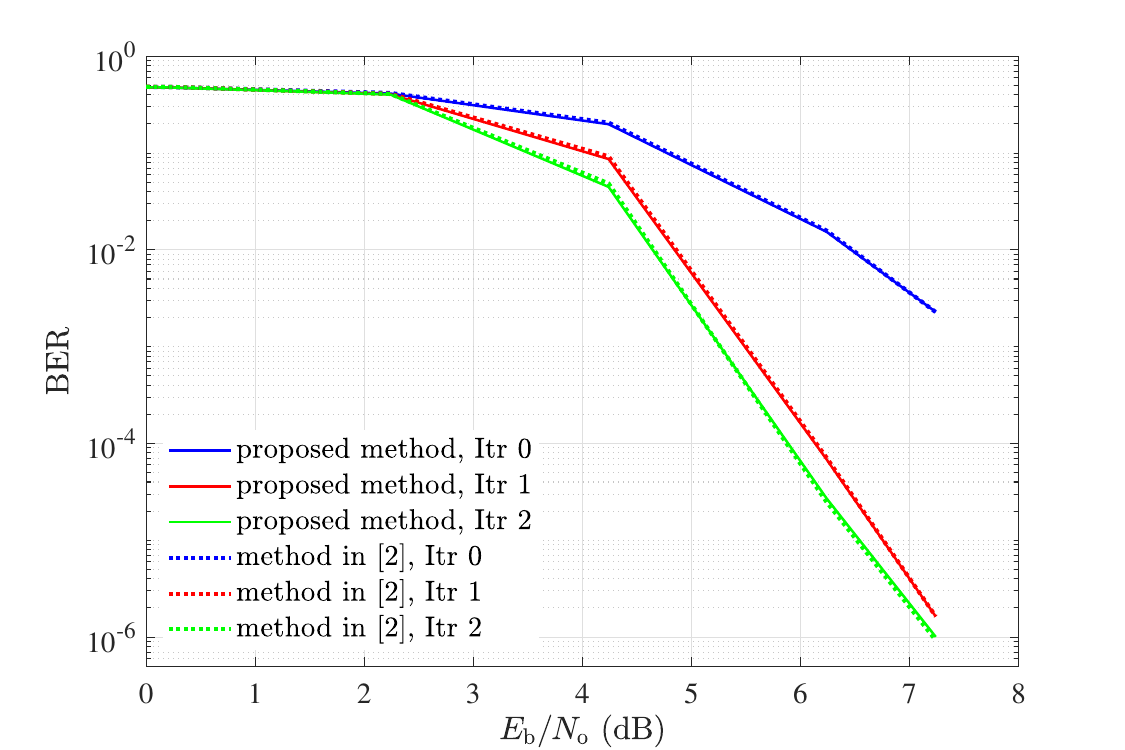}
\caption{{The BER performance of the proposed IM-based channel estimation algorithm for FTN signaling with $\tau = 0.84$ over \textit{channel model 2}.}}
\label{fig:BER_84}
\end{figure}

\begin{figure}[t]
\centering
\includegraphics[width=8.5cm,height=6cm]{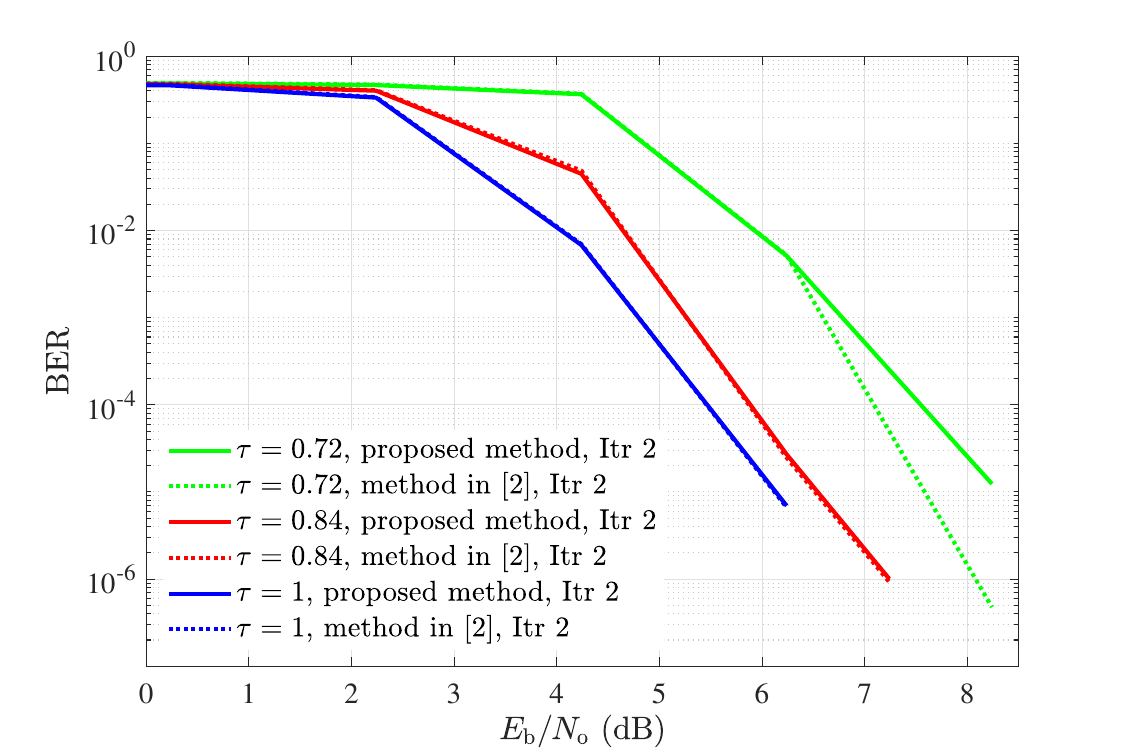}
\caption{{The BER performance of the proposed IM-based channel estimation algorithm for FTN signaling with $\tau = 0.72$, $\tau = 0.84$ and $\tau = 1$ over \textit{channel model 2}.}}
\label{fig:BER_compare}
\end{figure}

\begin{figure}[t]
\centering
\includegraphics[width=8.5cm,height=6cm]{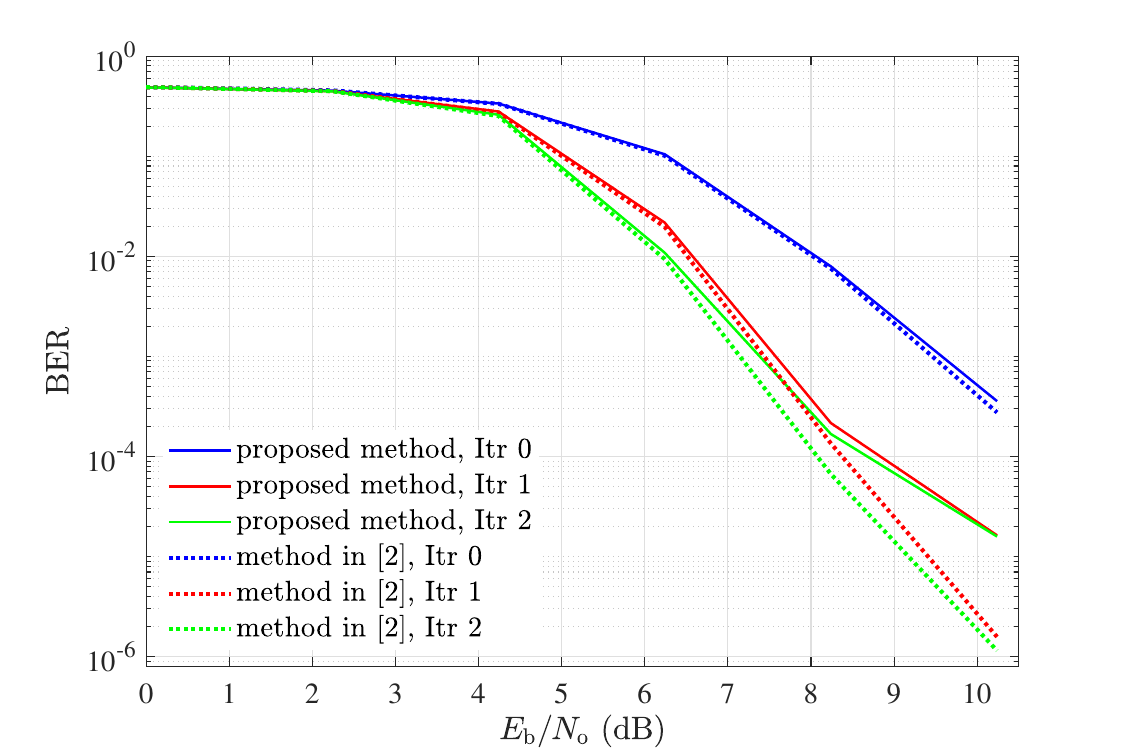}
\caption{{The BER performance of the proposed IM-based channel estimation algorithm for FTN signaling with $\tau = 0.84$ over \textit{channel model 1}.}}
\label{fig:BER_HF_84}
\end{figure}

\begin{figure}[t]
\centering
\includegraphics[width=8.5cm,height=6cm]{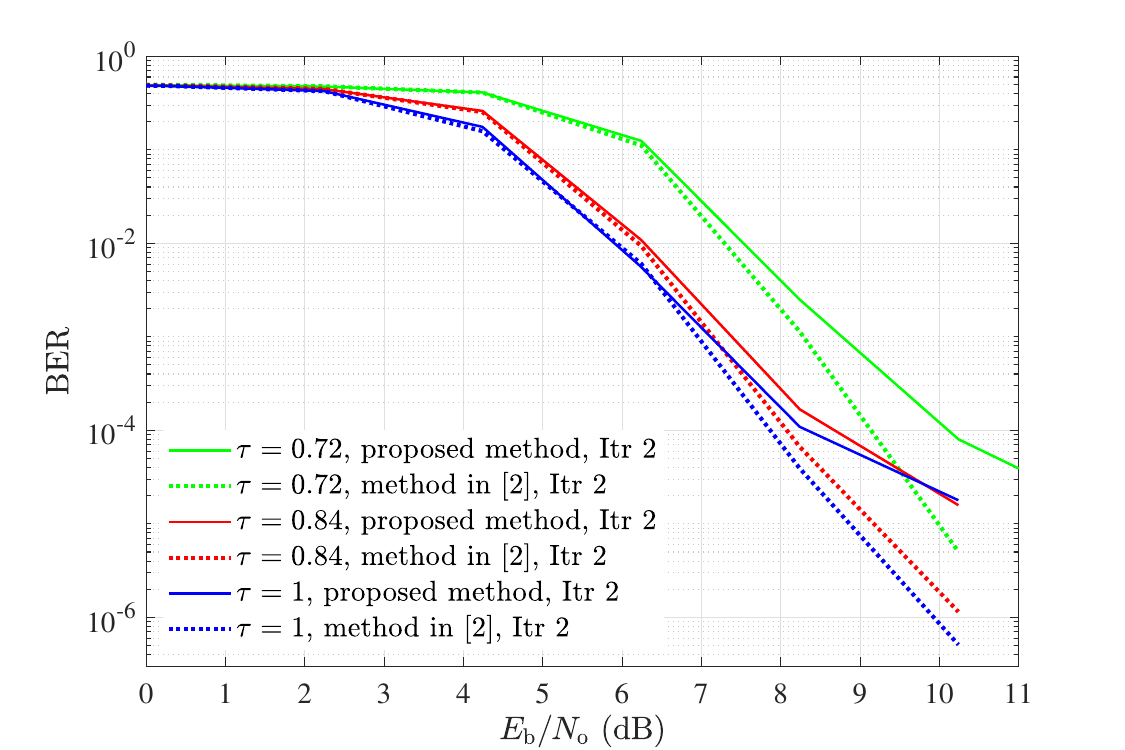}
\caption{{The BER performance of the proposed IM-based channel estimation algorithm for FTN signaling with $\tau = 0.72$, $\tau = 0.84$, and $\tau = 1$ over \textit{channel model 1}.}}
\label{fig:BER_HF_compare}
\end{figure}

\subsection{{BER Results}}
{
Fig. \ref{fig:BER_2017} illustrates a comparison of the BER performance of the proposed IM-based channel estimation algorithm for FTN signaling with $\tau = 0.8$ over \textit{channel model 3} with the BER of the work in \cite{ishihara2017iterative} considering $\beta = 0.5$. 
To ensure consistent overhead between the two studies, the pilot sequence length in the work in \cite{ishihara2017iterative} is configured to be 8 as discussed in detail in detail in subsection B of section ~\RomanNumeralCaps{5}.
In order to maintain a fair comparison, we implemented the channel estimation method in \cite{ishihara2017iterative} and measured the BER output of the decoder. 
Fig. \ref{fig:BER_2017} reveals that the proposed IM-based channel estimation algorithm
in this paper achieves a significantly better BER when compared to the BER resulting from the channel estimation approach in \cite{ishihara2017iterative}. 
After two iterations, the BER achieved with the proposed IM-based channel estimation algorithm exhibits about a 3.5 dB improvement compared to the case of \cite{ishihara2017iterative}.}

{
Fig. \ref{fig:BER_84} illustrates the BER performance of our proposed IM-based channel estimation algorithm designed for FTN signaling with $\tau = 0.84$ over \textit{channel model 2} with \cite{keykhosravi2023pilot}. Fig. \ref{fig:BER_84} shows that the BER of our proposed method remains the same as \cite{keykhosravi2023pilot}. As mentioned before, our proposed IM-based channel estimation algorithm for FTN signals has a $1.3\%$ SE gain compared to \cite{keykhosravi2023pilot}.}

{
Fig. \ref{fig:BER_compare} presents comparative results of the BER performance for our proposed IM-based channel estimation algorithm designed for FTN signaling with $\tau = 0.72$, $0.84$, and 1 over \textit{channel model 2} with \cite{keykhosravi2023pilot}. Fig. \ref{fig:BER_compare} shows that the BER of our proposed method remains the same as \cite{keykhosravi2023pilot}. 
As shown in Fig. \ref{fig:BER_compare} the BER of our proposed method remains the same as \cite{keykhosravi2023pilot} for the case of $\tau = 0.84$ and 1. Our proposed IM-based channel estimation algorithm for FTN signaling of $\tau = 0.84$ has a $1.3\%$ SE gain compared to \cite{keykhosravi2023pilot} and a $20.60\%$ SE gain compared to the Nyquist case.
For the case of $\tau = 0.72$, after two iterations, a degradation of 0.5 dB for $E_\text{b}/N_0$ of about 8 dB for the proposed IM-based channel estimation is observed when compared to the BER in \cite{keykhosravi2023pilot}. Therefore, the enhancement in SE in this scenario is accompanied by a degradation in the BER.}

{
Fig. \ref{fig:BER_HF_84} compares the BER performance of our proposed IM-based channel estimation algorithm designed for FTN signaling with $\tau = 0.84$ over \textit{channel model 1} with the method in \cite{keykhosravi2023pilot}. As shown in  Fig. \ref{fig:BER_HF_84}, the BER of our proposed method stays consistent with that of the method presented in \cite{keykhosravi2023pilot} for BER in the order of $10^{-4}$. However, there is about 1 dB degradation in the $E_\text{b}/N_0$ above 8 dB. This is because at high $E_\text{b}/N_0$ when the order of BER is less than $10^{-5}$, even a small PSLIE resulting in an error in the identification of pilot sequence location in a frame will lead to an error in data detection.}

{
Figure \ref{fig:BER_HF_compare} provides a comparative evaluation of the BER performance of our proposed IM-based channel estimation algorithm designed for FTN signaling. Different values of $\tau$ (0.72, 0.84, and 1) using \textit{channel model 1} are considered, and the comparison is made with the approach presented in \cite{keykhosravi2023pilot}.
As depicted in Fig. \ref{fig:BER_HF_compare}, our proposed method maintains a comparable BER to the approach outlined in \cite{keykhosravi2023pilot} for BER in the order of $10^{-3}$. Nevertheless, a degradation of approximately 1 dB is observed in the $E_\text{b}/N_0$ above 8 dB.
This occurs due to the impact of PSLIE becoming apparent at high $E_\text{b}/N_0$, particularly when BER order is less than $10^{-5}$.
}

\section{Conclusion}

In this paper, we proposed a novel IM-based channel estimation algorithm including PSP and PSLI algorithms to improve the SE for doubly-selective channel estimation in HF communications. The proposed PSP algorithm proficiently conveys additional information within the pilot sequence location of each frame for FTN signaling over frequency-selective channels. By harnessing the inherent characteristics of the HF channel and the advantageous auto-correlation properties of the optimal pilot sequence, we proposed the PSLI algorithm that effectively identifies pilot sequence locations within a given frame at the receiver. This is accomplished by demonstrating that the squared absolute value of the cross-correlation between the received symbols and the pilot sequence is a scaled representation of the squared absolute value of the auto-correlation of the pilot sequence, weighted by the gain of the corresponding HF channel path. {In this paper, we adopted an LSSE-based channel estimation approach for FTN signaling, aiming to estimate the complex channel coefficients at the pilot sequence locations within the frame. The proposed IM-based channel estimation algorithm of FTN signaling shows very low pilot sequence location identification error (less than $2 \times 10^{-6}$ and $10^{-3}$ for $E_\text{b}/N_0$ values higher than 6 dB and 12 dB for time-invariant and time-variant HF Poor channel, respectively, while considering $\tau=0.84$). Simulation results showed reaching similar performance in terms of the MSE and BER of the channel estimation compared to \cite{keykhosravi2023pilot} over frequency-selective channels while gaining SE improvement.} Simulation results demonstrated an enhancement in SE along with noteworthy improvement (about 6 dB) in the MSE of the channel estimation {as well as (about 3.5 dB) in BER of FTN signaling compared to the algorithm proposed in \cite{ishihara2017iterative} when dealing with frequency-selective channels. Simulation results for a higher order modulation (8-PSK modulation for both pilot sequence and data symbols) showed very low pilot sequence location identification error (less than $10^{-5}$ for $E_\text{b}/N_0$ higher than 6 dB for time-invariant HF Poor channel).}

\appendix
\label{app:cross}
\section*{Derivation of (\ref{equation34_2})}

In this appendix, we provide details on how we calculated the (\ref{equation34_2}). For simplification, we consider the transmission of one frame and drop the frame index $i$ over a general frequency selective time-invariant channel. Having the auto-correlation of $\tilde{p}_{k}$ in (\ref{equation31}), we could write the square of the absolute value of the $R_{\tilde{p}\tilde{p}} (\delta)$ for $\delta = 0, ... , N_p-1$ as

\begin{IEEEeqnarray}{rcl}\label{equationA1}
{|R_{\tilde{p}\tilde{p}} (\delta)|}^2 &{}={}&  
R_{\tilde{p}\tilde{p}}(\delta) \times {{R^\ast}_{\tilde{p}\tilde{p}}(\delta)} \nonumber\\
&{}={}& \sum \limits_{k = 0}^{N_p-1} \sum \limits_{k' = 0}^{N_p-1} { \tilde{p}_{k} \tilde{p^\ast}_{\delta+k} \tilde{p}_{k'} \tilde{p^\ast}_{\delta+k'} }.
\end{IEEEeqnarray}

Thus, for $\delta=0$, we can write ${|R_{\tilde{p}\tilde{p}} (0)|}^2$ as
\begin{IEEEeqnarray}{rcl}\label{equationA3}
{|R_{\tilde{p}\tilde{p}} (0)|}^2 &{}={}&  
 \sum \limits_{k = 0}^{N_p-1} \sum \limits_{k' = 0}^{N_p-1} { \tilde{p}_{k} \tilde{p^\ast}_{k} \tilde{p}_{k'} \tilde{p^\ast}_{k'} } \nonumber \\
 &{}={}&  \sum \limits_{k = 0}^{N_p-1} \sum \limits_{k' = 0}^{N_p-1}  {|\tilde{p}_{k}|}^2  {|\tilde{p}_{k'} |}^2.
\end{IEEEeqnarray}
We can write a vector of the output samples of the whitening filter in the (\ref{equation7}) %for a given frame $i$ and drop the frame index $i$ 
over a general frequency selective time-invariant channel for a sequence of $\tilde{\textbf{r}} = {[\tilde{r}_{0}, \tilde{r}_{1}, ..., \tilde{r}_{\text{N}-1}]}^\text{T}$ as
\begin{IEEEeqnarray}{rcl}\label{equationA5}
\tilde{\textbf{r}} &{}={}&  \textbf{s} \ast \textbf{c} \ast \textbf{v}+  \textbf{w}, \nonumber\\
%&{}={}&   \textbf{c} \ast \textbf{s} \ast \textbf{v}+  \textbf{w}\nonumber\\
&{}={}&   \textbf{c} \ast \tilde{\textbf{s}}+  \textbf{w},
\end{IEEEeqnarray}
where $\textbf{s} = {[s_0, s_1,..., s_{\text{N}-1}]}^\text{T}$ is a vector of data and pilot symbols in a transmitted frame, $\textbf{c}={[c_0, c_1,..., c_{L_\text{c}}]}^\text{T}$ is a vector of channel taps, $\textbf{v}={[v_0, v_1,..., v_{L_\text{h}-1}]}^\text{T}$ is the whitening filter, $\textbf{w}={[w_0, w_1,..., w_{\text{N}-1}]}^\text{T}$ is the AWGN noise vector and $\tilde{\textbf{s}} = \textbf{s} \ast \textbf{v}$.
Thus, the $k$th output sample of the whitening filter can be written as
\begin{IEEEeqnarray}{rcl}\label{equationA7}
\tilde{r}_{k} &{}={}&   \sum\limits_{l=0}^{L_\text{c}} c_{l} \tilde{s}_{k-l}+ w_k,  \quad  k = 0, ... , {{N_\text{s}}}-1 .
\end{IEEEeqnarray}
%where $\tilde{s}_k = \sum \limits_{m=0}^{\text{N}-1}   s_{m} v_{k-m}$.
Considering the received symbols in  (\ref{equationA7}) as a sequence of $\tilde{r}_{k}, k = 0,..., {{N_\text{s}}}-1$, the cross-correlation of $\tilde{r}_{k}$ and $\tilde{p}_{k}$ can be written as
\begin{IEEEeqnarray}{rcl}\label{equationA9}
R_{\tilde{r}\tilde{p}} (\delta) &{}={}&  \sum \limits_{m = 0}^{N_p-1} { \tilde{r}_{\delta+m} \tilde{p}_{m}^\ast},  \quad  \delta = 0, ... , {{N_\text{s}}}-1, \nonumber \\
&{}={}& \sum \limits_{m = 0}^{N_p-1}    \tilde{p}_{m}^\ast \left(\sum\limits_{l=0}^{L_\text{c}} c_{l}  {  \tilde{s}_{\delta+m-l} } \right) + \sum \limits_{m = 0}^{N_p-1} w_{\delta+m}  \tilde{p}_{m}^\ast, \nonumber \\
&{}={}& \sum\limits_{l=0}^{L_\text{c}} \sum \limits_{m = 0}^{N_p-1}    c_{l} \tilde{p}_{m}^\ast     {\tilde{s}_{\delta+m-l} }  + \sum \limits_{m = 0}^{N_p-1} w_{\delta+m}  \tilde{p}_{m}^\ast.
\IEEEeqnarraynumspace
\end{IEEEeqnarray}
Having the cross-correlation of $\tilde{r}_{k}$ and $\tilde{p}_{k}$ in (\ref{equationA9}), we could write the square of the absolute value of the $R_{\tilde{r}\tilde{p}} (\delta)$ as
\begin{IEEEeqnarray}{rcl}\label{equationA11}
{|R_{\tilde{r}\tilde{p}} (\delta)|}^2 &{}={}&  R_{\tilde{r}\tilde{p}}(\delta) \times {{R^\ast}_{\tilde{r}\tilde{p}}(\delta)} \nonumber\\
&{}={}& 
\sum\limits_{l'=0}^{L_\text{c}} \sum\limits_{l''=0}^{L_\text{c}} \sum \limits_{m = 0}^{N_p-1} \sum \limits_{m' = 0}^{N_p-1}   c_{l'} c_{l''}^\ast \tilde{p}_{m}^\ast  \tilde{p}_{m'}   {\tilde{s}_{\delta+m-l'} }  {\tilde{s}_{\delta+m'-l''}^\ast } \nonumber\\
& &{}+   
\sum \limits_{m = 0}^{N_p-1} \sum \limits_{m' = 0}^{N_p-1} w_{\delta+m} w_{\delta+m'}^\ast  \tilde{p}_{m}^\ast \tilde{p}_{m'} \nonumber\\
& &{}+
\sum\limits_{l'=0}^{L_\text{c}}  \sum \limits_{m = 0}^{N_p-1} \sum \limits_{m' = 0}^{N_p-1}   c_{l'}  \tilde{p}_{m}^\ast     {\tilde{s}_{\delta+m-l'} }  w_{\delta+m'}^\ast \tilde{p}_{m'}\nonumber\\
& &{}+
\sum\limits_{l''=0}^{L_\text{c}}  \sum \limits_{m' = 0}^{N_p-1} \sum \limits_{m = 0}^{N_p-1}   c_{l''}  \tilde{p}_{m'}^\ast     {\tilde{s}_{\delta+m'-l'} }  w_{\delta+m}^\ast \tilde{p}_{m}.
\
\IEEEeqnarraynumspace
\end{IEEEeqnarray}

For $\delta=n_{s_p}+l,  l=\{0, 1, ..., L_\text{c}\}$, we can rewrite the ${|R_{\tilde{r}\tilde{p}} (\delta)|}^2$ in (\ref{equationA11}) using the (\ref{equationA3}) as

\begin{IEEEeqnarray}{rcl}\label{equationA15}
{|R_{\tilde{r}\tilde{p}} (\delta=n_{s_p}+l)|}^2 &{}={}& 
 \sum \limits_{m = 0}^{N_p-1} \sum \limits_{m' = 0}^{N_p-1}   {|c_{l}|}^2   {|\tilde{p}_{m}|}^2  {|\tilde{p}_{m'} |}^2 + \zeta(\delta)\nonumber\\
&{}={}&
{|c_{l}|}^2 {|R_{\tilde{p}\tilde{p}} (0)|}^2+ \zeta(\delta).
\IEEEeqnarraynumspace
\end{IEEEeqnarray}

It should be noted that for the $\delta=n_{s_p}+l,  l=\{0, 1, ..., L_\text{c}\}$ and $l'=l$ and $l''=l$, we have $\tilde{s}_{\delta+m-l'} = \tilde{s}_{n_{s_p}+m} = \tilde{p}_{m}$ and $\tilde{s}_{\delta+m'-l''} = \tilde{s}_{n_{s_p}+m'} = \tilde{p}_{m'}$. It should be noted that $\zeta(\delta)$ is the interfered term and can be written as
\begin{IEEEeqnarray}{rcl}\label{equationA17}
\mkern-12mu \zeta(\delta) &={}&  \small{ \mkern-8mu \sum\limits_{\substack{l'=0 \\ l' \neq l}}^{L_\text{c}} \mkern-2mu \sum\limits_{\substack{l''=0 \\ l'' \neq l}}^{L_\text{c}} \mkern-5mu \sum \limits_{m = 0}^{N_p-1} \mkern-5mu \sum \limits_{m' = 0}^{N_p-1}  \mkern-8mu c_{l'} c_{l''}^\ast \tilde{p}_{m}^\ast  \tilde{p}_{m'}   {\tilde{s}_{n_{s_p}+m+l-l'} }  {\tilde{s}_{n_{s_p}+m'+l-l''}^\ast }\nonumber}\\
& &{}+
\sum \limits_{m = 0}^{N_p-1} \sum \limits_{m' = 0}^{N_p-1} w_{\delta+m} w_{\delta+m'}^\ast  \tilde{p}_{m}^\ast \tilde{p}_{m'} \nonumber\\
& &{}+
\sum\limits_{l'=0}^{L_\text{c}}  \sum \limits_{m = 0}^{N_p-1} \sum \limits_{m' = 0}^{N_p-1}   c_{l'}  \tilde{p}_{m}^\ast     {\tilde{s}_{\delta+m-l'} }  w_{\delta+m'}^\ast \tilde{p}_{m'}
\nonumber\\
& &{}+
\sum\limits_{l''=0}^{L_\text{c}}  \sum \limits_{m' = 0}^{N_p-1} \sum \limits_{m = 0}^{N_p-1}   c_{l''}  \tilde{p}_{m'}^\ast     {\tilde{s}_{\delta+m'-l'} }  w_{\delta+m}^\ast \tilde{p}_{m}.
\IEEEeqnarraynumspace
\end{IEEEeqnarray}

Thus, for all values of $\delta=n_{s_p}+l$ for which $l>L_\text{c}$, the $c_l=0$ and thus the first term in (\ref{equationA15}) is zero. We can write the ${|R_{\tilde{r}\tilde{p}} (\delta)|}^2$ as 
\begin{IEEEeqnarray}{rcl}\label{equationA18}
\small{|R_{\tilde{r}\tilde{p}} (\delta)|}^2 &=&  \nonumber \\
&&\mkern-12mu\small\begin{cases}
\small{|c_l|}^2 {|R_{\tilde{p}\tilde{p}} (0)|}^2 \!+\! \zeta(\delta),   \delta=n_{s_p}+l, l=\{0, 1, ..., L_\text{c}\},\!\vspace{12pt}\\
\small\zeta(\delta),\qquad \qquad \qquad \ \ \text{otherwise.}
\end{cases} \nonumber \\
\IEEEeqnarraynumspace
\end{IEEEeqnarray}
Thus, (\ref{equation34_2}) is proved.

\bibliographystyle{IEEEtran}
\bibliography{references}

\begin{thebibliography}{10}
\providecommand{\url}[1]{#1}
\csname url@samestyle\endcsname
\providecommand{\newblock}{\relax}
\providecommand{\bibinfo}[2]{#2}
\providecommand{\BIBentrySTDinterwordspacing}{\spaceskip=0pt\relax}
\providecommand{\BIBentryALTinterwordstretchfactor}{4}
\providecommand{\BIBentryALTinterwordspacing}{\spaceskip=\fontdimen2\font plus
\BIBentryALTinterwordstretchfactor\fontdimen3\font minus \fontdimen4\font\relax}
\providecommand{\BIBforeignlanguage}[2]{{%
\expandafter\ifx\csname l@#1\endcsname\relax
\typeout{** WARNING: IEEEtran.bst: No hyphenation pattern has been}%
\typeout{** loaded for the language `#1'. Using the pattern for}%
\typeout{** the default language instead.}%
\else
\language=\csname l@#1\endcsname
\fi
#2}}
\providecommand{\BIBdecl}{\relax}
\BIBdecl

\bibitem{ishihara2017iterative}
T.~Ishihara and S.~Sugiura, ``Iterative frequency-domain joint channel estimation and data detection of faster-than-{N}yquist signaling,'' \emph{IEEE Transactions on Wireless Communications}, vol.~16, no.~9, pp. 6221--6231, Sep. 2017.

\bibitem{keykhosravi2023pilot}
S.~Keykhosravi and E.~Bedeer, ``Pilot design and doubly-selective channel estimation for faster-than-{N}yquist signaling,'' \emph{Physical Communication}, pp. 102--122, Jun. 2023.

\bibitem{wang2018hf}
J.~Wang, G.~Ding, and H.~Wang, ``H{F} communications: Past, present, and future,'' \emph{China Communications}, vol.~15, no.~9, pp. 1--9, Sep. 2018.

\bibitem{muresan2015ionospheric}
O.~A. Muresan, A.~Pastrav, E.~Puschita, and T.~Palade, ``Ionospheric {HF} channel modeling and end-to-end {HF} system simulation.'' \emph{Acta Technica Napocensis. Electronica-Telecomunicatii}, vol.~56, no.~4, Nov. 2015.

\bibitem{navarro2004error}
A.~Navarro, R.~Rodrigues, J.~Angeja, J.~Tavares, L.~Carvalho, and F.~Perdigao, ``An error-resilient approach for real-time packet communications by {HF}-channel diversity,'' in \emph{Digital Wireless Communications}, vol.~VI.\hskip 1em plus 0.5em minus 0.4em\relax SPIE 5440, 2004, pp. 67--76.

\bibitem{anderson2013faster}
J.~B. Anderson, F.~Rusek, and V.~\"{O}wall, ``faster-than-{N}yquist signaling,'' \emph{Proceedings IEEE}, vol. 101, no.~8, pp. 1817--1830, Aug. 2013.

\bibitem{li2020beyond}
Q.~Li, F.-K. Gong, P.-Y. Song, G.~Li, and S.-H. Zhai, ``Beyond {DVB-S2X}: faster-than-{N}yquist signaling with linear precoding,'' \emph{IEEE Transactions on Broadcasting}, vol.~66, no.~3, pp. 620--629, Sep. 2020.

\bibitem{john2008digital}
J.~G. Proakis and M.~Salehi, \emph{Digital communications}.\hskip 1em plus 0.5em minus 0.4em\relax McGraw-Hill., 2008.

\bibitem{liveris2003exploiting}
A.~D. Liveris and C.~N. Georghiades, ``Exploiting faster-than-{N}yquist signaling,'' \emph{IEEE Transactions on Communications,}, vol.~51, no.~9, pp. 1502--1511, Sep. 2003.

\bibitem{prlja2008receivers}
A.~Prlja, J.~B. Anderson, and F.~Rusek, ``Receivers for faster-than-{N}yquist signaling with and without turbo equalization,'' in \emph{Proceedings IEEE International Symposium on Information Theory (ISIT)}, 2008, pp. 464--468.

\bibitem{bedeer2017reduced}
E.~Bedeer, H.~Yanikomeroglu, and M.~H. Ahmed, ``Reduced complexity optimal detection of binary faster-than-{N}yquist signaling,'' in \emph{Proceedings IEEE International Conference on Communications (ICC)}, May 2017, pp. 1--6.

\bibitem{hong2016performance}
S.~B. Hong and J.-S. Seo, ``Performance enhancement of {F}aster-{T}han-{N}yquist signaling based single-carrier frequency-domain equalization systems,'' \emph{IEEE Transactions on Broadcasting}, vol.~62, no.~4, pp. 918--935, Nov. 2016.

\bibitem{sugiura2013frequency}
S.~Sugiura, ``Frequency-domain equalization of faster-than-{N}yquist signaling,'' \emph{IEEE Wireless Communications Letters}, vol.~2, no.~5, pp. 555--558, Oct. 2013.

\bibitem{shi2017frequency}
Q.~Shi, N.~Wu, X.~Ma, and H.~Wang, ``Frequency-domain joint channel estimation and decoding for faster-than-{N}yquist signaling,'' \emph{IEEE Transactions on Communications}, vol.~66, no.~2, pp. 781--795, Oct. 2017.

\bibitem{bedeer2017very}
E.~Bedeer, M.~H. Ahmed, and H.~Yanikomeroglu, ``A very low complexity successive symbol-by-symbol sequence estimator for faster-than-{N}yquist signaling,'' \emph{IEEE Access}, vol.~5, pp. 7414--7422, Mar. 2017.

\bibitem{ibrahim2021novel}
A.~Ibrahim, E.~Bedeer, and H.~Yanikomeroglu, ``A novel low complexity faster-than-{N}yquist ({FTN}) signaling detector for ultra high-order {QAM},'' \emph{IEEE Open Journal of the Communications Society}, vol.~2, pp. 2566--2580, Nov. 2021.

\bibitem{hirano2014tdm}
T.~Hirano, Y.~Kakishima, and M.~Sawahashi, ``T{DM} based reference signal multiplexing for faster-than-{N}yquist signaling using {OFDM/OQAM},'' in \emph{Proceeding IEEE International Conference on Communication Systems (ICCS)}, Nov. 2014, pp. 437--441.

\bibitem{wu2017hybrid}
N.~Wu, W.~Yuan, Q.~Guo, and J.~Kuang, ``A hybrid {BP-EP-VMP} approach to joint channel estimation and decoding for {FTN} signaling over frequency selective fading channels,'' \emph{IEEE Access}, vol.~5, pp. 6849--6858, May. 2017.

\bibitem{mao2018novel}
T.~Mao, Q.~Wang, Z.~Wang, and S.~Chen, ``Novel index modulation techniques: A survey,'' \emph{IEEE Communications Surveys \& Tutorials}, vol.~21, no.~1, pp. 315--348, Firstquarter, 2018.

\bibitem{ma2020joint}
Y.~Ma, N.~Wu, W.~Yuan, D.~W.~K. Ng, and L.~Hanzo, ``Joint channel estimation and equalization for index-modulated spectrally efficient frequency division multiplexing systems,'' \emph{IEEE Transactions on Communications}, vol.~68, no.~10, pp. 6230--6244, Oct. 2020.

\bibitem{cavers2006mobile}
J.~Cavers, \emph{Mobile channel characteristics}.\hskip 1em plus 0.5em minus 0.4em\relax Springer Science \& Business Media, 2006, vol. 555.

\bibitem{ITU}
\emph{ITU-R F.520: Use of high frequency ionospheric channel simulators}, 1992.

\bibitem{ITUF}
\emph{IITU-R F.1487: Testing of HF modems with bandwidths of up to about 12 KHz using ionospheric channel simulators}, 2000.

\bibitem{otnes2002improved}
R.~Otnes, ``Improved receivers for digital high frequency communications: Iterative channel estimation, equalization, and decoding (adaptive turbo equalization),'' 2002.

\bibitem{wright1997primal}
S.~J. Wright, \emph{Primal-dual interior-point methods}.\hskip 1em plus 0.5em minus 0.4em\relax SIAM, 1997.

\end{thebibliography}
%\end{multicols} %just before \end{document}

\end{document}